\documentclass[onefignum,onetabnum]{siamonline190516}

\usepackage{lipsum}
\usepackage{amsfonts}
\usepackage{graphicx}
\usepackage{epstopdf}
\usepackage{subcaption}
\usepackage{algorithmic}

\pdfoutput=1

  \DeclareGraphicsExtensions{.eps,.pdf,.png,.jpg}

\usepackage{tikz}
\usetikzlibrary{shapes,arrows, shapes.geometric,calc, snakes}
\tikzstyle{decision} = [diamond, draw, fill=gray!20, 
text width=8em, text badly centered, node distance=4cm, inner sep=0pt]
\tikzstyle{block} = [rectangle, draw, fill=gray!20, 
text width=5em, text centered, rounded corners, minimum height=4em, node distance=4cm]
\tikzstyle{line} = [draw, -latex']
\tikzstyle{cloud} = [draw, ellipse,fill=gray!20, text width= 9em, text centered, node distance=7cm,
minimum height=4em]

\pgfmathsetmacro{\radius}{0.25}
\pgfmathsetmacro{\length}{5}

\tikzset {
	bent cylinder/.pic={
		\pgfmathsetmacro{\anglebottom}{90-20+40*rnd}
		\pgfmathsetmacro{\strengthbottom}{\length*(0.3+rnd/3)}
		\pgfmathsetmacro{\angletop}{90-20+40*rnd}
		\pgfmathsetmacro{\strengthtop}{\length*(0.3+rnd/3)}
		
		\fill[fill=white] 
		(\radius,0) 
		.. controls +(\anglebottom:\strengthbottom) 
		and +(-\angletop:\strengthtop) 
		.. (\radius,\length) 
		arc (360:180:\radius cm and 0.5*\radius cm) 
		.. controls +(-\angletop:\strengthtop) 
		and +(\anglebottom:\strengthbottom) 
		.. (-\radius,0) 
		;
		
		\draw[gray!90, ultra thick, dashed, line join=round]
		(\radius,0) 
		.. controls +(\anglebottom:\strengthbottom) 
		and +(-\angletop:\strengthtop) 
		.. (\radius,\length) 
		arc (360:180:\radius cm and 0.5*\radius cm) 
		.. controls +(-\angletop:\strengthtop) 
		and +(\anglebottom:\strengthbottom) 
		.. (-\radius,0) 
		;
		
		\draw[gray!90, ultra thick, line join=round]
		(0,0) 
		.. controls +(\anglebottom:\strengthbottom) 
		and +(-\angletop:\strengthtop) 
		.. (0,\length) 
		;
		
		\filldraw[ultra thick, dashed, draw=gray!90, fill=white]
		(0,\length) circle (\radius cm and 0.5*\radius cm);
		
		\draw[ultra thick, dashed, draw=gray!90]
		(0,0) circle (\radius cm and 0.5*\radius cm);
	}
}

\usepackage{enumitem}
\setlist[enumerate]{leftmargin=.5in}
\setlist[itemize]{leftmargin=.5in}

\newsiamremark{remark}{Remark}
\newsiamremark{hypothesis}{Hypothesis}
\newsiamremark{problem}{Problem}
\crefname{hypothesis}{Hypothesis}{Hypotheses}
\newsiamthm{claim}{Claim}

\headers{One-way FBI}{N. Hagmeyer, M. Mayr, I. Steinbrecher, and A. Popp}

\title{Fluid-beam interaction: Capturing the effect of embedded slender bodies on global fluid flow and vice versa\thanks{Submitted to the editors DATE.}}

\author{Nora Hagmeyer\thanks{Institute for Mathematics and Computer-Based Simulation, University of the Bundeswehr Munich, Werner-Heisenberg-Weg 39, 85577 Neubiberg, Germany 
  (\email{nora.hagmeyer@unibw.de}, \url{unibw.de/imcs-en}).}
\and Matthias Mayr\footnotemark[2]\ \thanks{Data Science \& Computing Lab, University of the Bundeswehr Munich, Werner-Heisenberg-Weg 39, 85577 Neubiberg, Germany}
\and Ivo Steinbrecher\footnotemark[2]
\and Alexander Popp\footnotemark[2]}

\usepackage{amsopn}

\graphicspath{{pictures/mixer/}{pictures/hindernis_lowre/}{pictures/stent/}{pictures/beam_flow/}{pictures/flow_oscillating/}{pictures/contact/}}

\ifpdf
\hypersetup{
  pdftitle={One-way FBI},
  pdfauthor={N.Hagmeyer, M. Mayr, I. Steinbrecher, and A. Popp}
}
\fi

\begin{document}

\maketitle

\begin{abstract}
This work addresses research questions arising from the application of geometrically exact beam theory in the context of fluid-structure interaction (FSI). Geometrically exact beam theory has proven to be a computationally efficient way to model the behavior of slender structures while leading to rather well-posed problem descriptions. In particular, we propose a mixed-dimensional embedded finite element approach for the coupling of one-dimensional geometrically exact beam equations to a three-dimensional background fluid mesh, referred to as fluid-beam interaction (FBI) in analogy to the well-established notion of FSI. Here, the fluid is described by the incompressible isothermal Navier-Stokes equations for Newtonian fluids. In particular, we present algorithmic aspects regarding the solution of the resulting one-way coupling schemes and, through selected numerical examples, analyze their spatial convergence behavior as well as their suitability not only as stand-alone methods but also for an extension to a full two-way coupling scheme.
\end{abstract}



\section{Introduction}
\label{sec:intro}

The interaction of rod-like structures with fluid flow plays an important role in a broad spectrum of applications varying from biomechanical to industrial processes. Examples include the interaction of endovascular devices with blood flow \cite{Gay2005}, coiling during the treatment of cerebral aneurysms \cite{mitsos2008}, the effect of submerged vegetation on surrounding fluid \cite{oconnor2019, wang2019}, the application of brush seals as well as the use of fibrous coverings in flow control \cite{favier2009, kunze2012}, to name a few. The numerical modeling of problems involving such rod-like structures  with classical continuum-based finite elements poses a challenge because it promptly leads to locking effects as well as very large system sizes. In order to ensure a well-posed formulation as well as an efficient usage of computational resources, applications can often benefit from employing one-dimensional (1D) beam theory to model the slender bodies compared to a full three-dimensional (3D) solid formulation.

The application of dimensionally reduced models to adequately describe complex behavior of slender bodies has a long-standing history. The formulations in this paper greatly build on the extension of such models to the nonlinear regime of finite deformations in three dimensions. Important to mention in this context is the extension to a finite strain problem in 3D by Simo in \cite{simo1985} based on the work by Reissner in \cite{Reissner1972}. The formulations proposed in these works fall in the group of geometrically exact beam theories and constitute the basis of computationally efficient and rather well-posed problems for systems of slender bodies. The actual beam implementations used in the following constitute an extension and application of the above ideas by Meier et al. in \cite{Meier2014, Meier2015, meier_contact}.

These reduced-dimensional formulations can be used to model highly complex systems of slender struts. In \cite{tambavca2010mathematical}, Tamba\v{c}a et al. propose a general one-dimensional vascular stent model, which is utilized in \cite{Tambaca2010} to efficiently simulate the behavior of coronary stents under physiologically reasonable conditions. As an extension of that, the behavior of this one-dimensional stent model is compared to the results of a fully resolved three-dimensional simulation in \cite{zunino2016}. A difference of less than $6\%$ in the displacement magnitude at a complexity reduction, in terms of degrees of freedom (DoFs), of more than 400 was observed.

Staying with the application of stenting procedures for the moment, it becomes clear that not only the behavior of the stent but especially its interaction with the arterial wall, be it chemical in the case of drug-eluding stents, or mechanical in nature, is of interest. Additionally, "[a]n extension of [such a] model to account for flow perturbations would be beneficial and is in order"\cite[p. 614]{zunino2016}, in particular to model growth and restenosis as a response to oscillating wall shear stresses (WSS). In general, this necessitates approaches for the interaction between the 1D beam model with classical continuum-based equations describing the blood flow and the mechanics of the vessel wall.

Such mixed-dimensional models are not restricted to endovascular devices, but naturally arise in a multitude of applications including structures with high slenderness ratios. The solution of these mixed-dimensional interaction problems leads to new challenges compared to equal-dimensional problems. The most prominent among them is the question of how to couple three-dimensional variables with 1D stress resultants as well as the challenge of transferring values between unavoidably geometrically non-matching meshes.

In the area of fluid-structure interaction (FSI), the immersed boundary method (IBM) \cite{PESKIN1972252, Peskin2002, Peskin2006} constitutes a well-studied method for mixed-dimensional interactions. Nearly five decades ago, Peskin first explored the idea of embedding a one-dimensional body representing the response of a thin structure within a two-dimensional fluid domain in order to simulate the blood flow in the vicinity of a heart valve \cite{PESKIN1972252}. At the time, Peskin modeled the blood flow via the incompressible Navier-Stokes equations spatially discretized with a finite difference method (FD). The force exerted by the heart valve itself on the fluid around it was modeled as a direct linear response to the fluid velocity. The interpolation of the fluid velocity and force response between the two meshes was realized by using so-called delta functions as numerical approximations of Dirac functions.

Since that time a significant amount of research was conducted on the extension and application of the IBM to fully-resolved equal-dimensional 2D and 3D FSI problems based on fictitious domain methods, cf. \cite{Rauch2018, hesch2014, Baaijens2001, IFEM2007, Gay2005} among many others. In such methods, the domain occupied by the immersed structure is filled by a fictitious fluid volume, which usually leads to a coupling of the involved FSI values on the fictitious domain volume and/or its surface. An exception represents the work of Baaijens in \cite{Baaijens2001}, in which applications of his proposed fictitious domain/mortar finite element methods to slender bodies are shown. The slender bodies are modeled with continuum-based equations, while the coupling interface is reduced to only one side of the fictitious domain. Comparisons with fully coupled Arbitrary Langrangian Eulerian (ALE) based methods show the validity of this simplification.

In contrast, one-dimensional beam formulations are embedded in a two-dimensional fluid in \cite{Henshaw2015, wang2019}. In \cite{Henshaw2015}, the coupling of the fluid and the beam is realized with moving composite meshes interacting on the beam surface, while an IBM-type coupling of the beam surface with the background mesh is presented in \cite{wang2019}. Finally, Huang et al. \cite{huang2019} present results for an IBM-type method coupling a one-dimensional beam formulation with a three-dimensional fluid. Again, the beam is coupled with the Navier-Stokes equations on the reconstructed beam surface, making all of the mentioned approaches surface-to-volume coupling schemes as categorized in \cite{steinbrecher2020}. Within these approaches, the slender body is modeled using one-dimensional equations, the fluid domain consists of the entire simulation domain, and the fluid-beam interaction quantities are coupled on the reconstructed beam surface. This makes it necessary to reconstruct the beam surfaces to be coupled with the fluid equations, and in turn, to exchange relevant data between the beam surface and its centerline. As a result, the fluid domain has to be massively refined in order to resolve the surface mesh tying, and also the coupling procedure itself rapidly grows more complex in terms of computational efficiency, somewhat reducing the computational advantages of employing a reduced order beam model. The work of Tschisgale et al. \cite{Tschisgale2020} represents an intermediate step, since the beam surface does not have to be reconstructed. Instead, classical regularized delta functions, depending on the beam radius, are used to add fluid-beam interaction forces to the overall problem. This method, nevertheless, also necessitates the use of finely resolved background meshes in order for the delta functions to have a width of multiple fluid elements. A truly mixed-dimensional fluid-beam interaction method, that allows for relatively coarse background meshes, was discussed only in \cite{wu2019}. Here, the coupling was realized on the beam centerline and applied to the simulation of a transcatheter heart valve. The computational results of the biomechanical problem look promising, however, no numerical study of the necessary assumptions for the validity of the method nor its limitations were presented.

Within this contribution, we now propose an immersed mixed-dimensional FSI formulation, embedding slender bodies modeled by geometrically exact beam theory within a three-dimensional incompressible fluid flow. Both, the beam formulation and the employed incompressible isothermal Navier-Stokes equations, are discretized with the finite element method (FEM). The proposed formulation is based on the assumption that the beam diameter is smaller than the characteristic fluid element diameter, and that the development of global flow phenomena, in contrast to interface phenomena, is of primary interest. With such application cases in mind, we will follow the simplification ideas of Baaijens \cite{Baaijens2001, wu2019} and directly couple the one-dimensional beam centerline to the three-dimensional continuum equations following a Gauss-point-to-segment (GPTS) type approach, e.g. as commonly used in contact mechanics.
This work constitutes the first rigorous numerical study of a truly mixed-dimensional 1D-3D coupling approach to capture global effects of the interaction of nonlinear beam elements with 3D fluid flow in a manner that enables the use of a relatively coarse background mesh compared to the diameter of the slender beam. Our method, thus, allows for a very efficient solution of the FBI problem in terms of the required number of unknowns. For highly slender embedded fibers, the use of a relatively coarse fluid mesh is crucial in order to conserve benefits of using a reduced-dimensional structure formulation in terms of the system size, solver time as well as the evaluation of the coupling terms itself. These advantages become even more pronounced for applications including a large number of embedded fibers.
Similar coupling approaches have already shown promising results for solid-beam interaction in our previous work \cite{steinbrecher2020} and for the modeling of tumor growth in \cite{kremheller2019}.

The remaining part of this paper will be structured as follows: In Section \ref{sec:equations}, we will present the single field equations as well as the continuous FBI problem. Section \ref{sec:numerics} will briefly outline the discretization methods used to treat the partial differential equations (PDEs), the load and motion transfer schemes used to interpolate between the two domains as well as the numerical integration of the coupling terms. In Section \ref{sec:algo}, we will introduce one-way coupling algorithms used for the subsequent numerical results. In Section \ref{sec:examples}, we will proceed to demonstrate the applicability of the proposed approach by validating results of the one-way coupling schemes in the two special cases of (i) very light fibers as well as (ii) almost rigid beam structures, by discussing numerical features of the proposed approach as well as studying its behavior under uniform mesh refinement with respect to a 3D reference solution.

\section{Problem Formulation}
\label{sec:equations}

In this section we briefly recount the governing equations for the fluid field, namely the incompressible isothermal Navier-Stokes equations for Newtonian fluids, as well as the geometrically exact beam formulations applied to the numerical examples in Section \ref{sec:examples}. Both fields are coupled via an IBM-type approach applied on the beam centerline in the current configuration as detailed in Section \ref{subsec:fbi}. For improved readability, the superscript $\left(\cdot\right)^f$ shall denote fluid-related quantities. For the beam equations, we will use the abbreviation $\left(\dot{x}\right):=\dfrac{\partial x}{\partial t}$ for the time derivative of $x$, and $\left({x}^\prime\right):=\dfrac{\partial x}{\partial s}$ for the derivative with respect to the curve parameter $s$ as is common practice in beam theory. In turn, we will abstain from a rigorous use of a superscript to denote all beam-related quantities. Nevertheless, the superscript $\left(\cdot\right)^b$ will be used to highlight beam-related quantities where appropriate.

   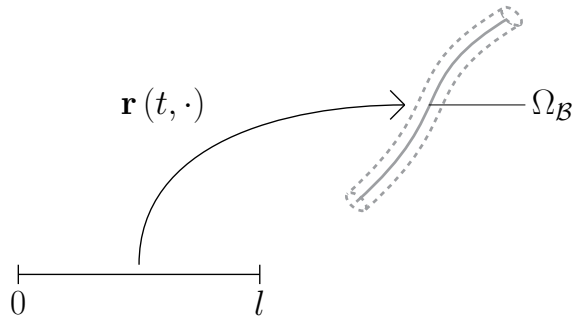
\begin{figure}[h!]
	\begin{center}
		\resizebox{0.5\textwidth}{!}{
			\begin{tikzpicture}
			\draw[thick] (-7, 0)  -- (-2, 0);
			\draw[thick] (-7, 0.2)  -- (-7, -0.2) node[below] {\huge{$0$}};
			\draw[thick] (-2, 0.2)  -- (-2, -0.2) node[below] {\huge{$l$}};
			\draw[thick] (-4.5, 0.2) to[out=90,in=180] (1,3.5);
			\draw[thick] (1,3.5) to (0.7, 3.8);
			\draw[thick] (1,3.5) to (0.7, 3.2);
			\node at (-4, 3.5) {\huge{$\mathbf{r}\left(t, \cdot\right)$}};
			\pgfmathsetseed{2030}
			\draw (0, 3/2) pic[rotate=-40] {bent cylinder};
			\draw (1.5,3.5) -- (3.5,3.5) node[anchor=west] {\huge{$\Omega_\mathcal{B}$}};
			\end{tikzpicture}
		}
		\caption{Depiction of a beam and its centerline representation by a curve}
	\end{center}
\end{figure}

\subsection{Geometrically Exact Beam Theory}
\label{subsec:beams}

We will use geometrically exact beam models in the sense that "the relationships between the configuration and the strain measures are consistent with the virtual work principle and the equilibrium equations at a deformed state regardless of the magnitude of displacements, rotations and strain"\cite[p. 1126]{crisfield1999}. The two geometrically exact beam formulations treated within this work fall into the category of semi-induced beam theories, meaning that the kinematic equations are consistently derived from the three-dimensional continuum theory while the constitutive equations are directly postulated on the one-dimensional geometry. In contrast to fully induced methods this approach ensures the fulfillment of important mechanical principles such as the equilibrium of forces \cite{Meier2014}.

In order to write the two particular beam formulations in the upcoming subsections, namely the Simo-Reissner (SR) model and a torsionfree (TF) variant,  in a compact form, let
\begin{equation}
\mathcal{V^\text{p}_B}:=\left\{v: \left[0,l\right] \rightarrow \mathbb{R}^3:\left\|\dfrac{\partial^\alpha v}{\partial s}\right\|_\mathcal{B}<\infty, \alpha \leq p \right\} \qquad \text{for} \quad p \in \left\{1,2\right\},
\end{equation} be two different spaces of curves on $\left[0,l\right]$ with additional requirements on the smoothness of higher order derivatives of the contained curves. Both spaces shall be endowed with the inner product $\left(\cdot , \cdot\right)_\mathcal{B}:=\int\limits_\mathbb{R}\left(\cdot , \cdot \right)_{\mathbb{R}^3}\text{ d}s$ and the associated norm $\left\|\cdot\right\|_\mathcal{B}:=\sqrt{\left(\cdot , \cdot\right)_\mathcal{B}}$. Here, $\left(\cdot , \cdot \right)_{\mathbb{R}^3}$ simply denotes the scalar product of two vectors in 3D and $\left\|\cdot\right\|_{\mathbb{R}^3}:=\sqrt{\left(\cdot , \cdot \right)_{\mathbb{R}^3}}$ its associated norm.
Furthermore, in the upcoming subsections, we will denote the cross product of two vectors $\mathbf a$ and $\mathbf b$ using the cross product operator $\boldsymbol{S}\left(\boldsymbol{a}\right)\boldsymbol{b}:=\mathbf{a} \times \mathbf{b}$.

In general, the state of a beam at each time $t\in I:=\left[0,T\right]$ can be characterized by the position of its centerline $\boldsymbol{r}\left(t, \cdot\right)\in \mathcal{V}^\text{p}_r := \mathcal{V^\text{p}_B}$, for $p \in\left\{1,2\right\}$, and the cross-section rotation vector $\boldsymbol{\psi}\left(t, \cdot\right)\in \mathcal{V}_\theta:=\mathcal{V^\text{1}_B}$. With the help of Rodrigues' rotation formula, the rotation vector can be used to define a rotation matrix $\boldsymbol{\Lambda}=\boldsymbol{\Lambda}\left(\boldsymbol{\psi}\left(t,s\right)\right)\in SO\left(3\right)$ in the special orthogonal group $SO\left(3\right)$ for all $s\in \left[0,l\right]$, $t\in I$.

To maintain a compact notation where possible, we introduce the solution state $\boldsymbol{\eta}^b\in \mathcal{V_B}$ summarizing all relevant unknowns, such as the centerline position $\mathbf{r}$ and the rotation vector $\boldsymbol{\psi}$, to represent the state of the regarded beam. The exact form of $\boldsymbol{\eta}^b$ and $\mathcal{V_B}$ depends on basic modeling assumptions and therefore differs for the two beam models, which will be introduced in the upcoming subsections.

For now, using the presented general notation, we can complete the general beam problem by introducing the Dirichlet boundary conditions $\boldsymbol{\eta}_D$ and an initial state $\boldsymbol{\eta}_0$. Then, the state of the beam can be characterized as solution $\boldsymbol{\eta}^b$ of the following problem:

\begin{problem}
	\label{prob:beam}
	Find $\boldsymbol{\eta}^b\in L^2\left(I, \mathcal{V_B}+\boldsymbol{\eta}_D\right)$ such that
	
	\begin{equation}
	\int\limits_0^T \mathbf{a}^b\left(\boldsymbol{\eta}^b; \delta\boldsymbol{\eta}^b\right) - \mathbf{b}^b\left( \delta\boldsymbol{\eta}^b\right) \text{ d}t=0 \qquad \forall \quad \delta\boldsymbol{\eta}^b\in \mathcal{V_B},
	\end{equation}
	with $\boldsymbol{\eta}^b=\boldsymbol{\eta}_0$ a. e. for $t=0$, and $\left\|\mathbf{r}^\prime\left(0, \cdot\right)\right\|_{\mathbb{R}^3}=1$ a. e. on $\left[0,l\right]$.
\end{problem}
The exact form of the semi-linear form $\boldsymbol{a}^b\left(\boldsymbol{\eta}^b;\delta\boldsymbol{\eta}^b\right)$, the linear form $\boldsymbol{b}^b\left(\delta\boldsymbol{\eta}^b\right)$, and the corresponding solution space $\mathcal{V_B}$ will be given in the upcoming subsections for two particular beam models.

\begin{remark}
In problem \ref{prob:beam}, the additional condition $\left\|\mathbf{r}^\prime\left(0, \cdot\right)\right\|_{\mathbb{R}^3}=1$ means that $\mathbf{r}\left(0, \cdot\right)$ represents an arc-length parameterized curve on $[0,l]$, or, in other words, $l$ defines the initial length of the beam.
\end{remark}
 
 \subsubsection{Simo-Reissner Beam Model}
 The Simo-Reissner (SR) model is based on the assumptions of plane, rigid cross-sections, but does not introduce any additional kinematic constraints on the beam. For further details on the derivation, see e.g. \cite{simo1985}.
 
 For the density $\rho^b$, shear and Young's modulus $G$ and $E$, respectively, the cross-section area $A$, the reduced cross-section values $A_1$ and $A_2$, the torsional moment of inertia $I_T$, and the principal moments of inertia $I_2$, $I_3$, the constitutive matrices $\boldsymbol{C}_M$, $\boldsymbol{C}_F$, and the inertia tensor $\boldsymbol{C}_\rho$ are given as follows:
 
 \begin{eqnarray*}
 \boldsymbol{C}_M:=\left(\begin{array}{ccc}
 GI_T & 0 & 0 \\
 0 & EI_2 & 0 \\
 0 & 0 & EI_3
 \end{array}
 \right),
  \boldsymbol{C}_F:=\left(\begin{array}{ccc}
 EA & 0 & 0 \\
 0 & GA_2 & 0 \\
 0 & 0 & GA_3
 \end{array}
 \right),
  \boldsymbol{C}_\rho:=\left(\begin{array}{ccc}
 	\rho^b\left(I_2+I_3\right) & 0 & 0 \\
 	0 & \rho^bI_2 & 0 \\
 	0 & 0 & \rho^bI_3
 \end{array}
 \right).
 \end{eqnarray*}
 
 With the deformation measures $\boldsymbol{\Gamma}:=\boldsymbol{\Lambda}^T\mathbf{r}^\prime-\mathbf{e}_1$, and the solution state $\boldsymbol{\eta}^b:=\left(\mathbf{r}, \boldsymbol{\psi}\right)$, the semi-linear form in Problem \ref{prob:beam} is defined as
 
\begin{equation}
\begin{split}
\mathbf{a}^b\left(\boldsymbol{\eta}^b; \delta\boldsymbol{\eta}^b\right) &:= \left(\rho^b \mathbf{A\mathbf{\ddot{r}}}, \delta\mathbf{r}\right)_\mathcal{B} + \left(\boldsymbol{\Lambda} \mathbf{C}_F \boldsymbol{\Gamma}, \delta\mathbf{r}^\prime \right)_\mathcal{B} \\
&+ \left(\boldsymbol{\Lambda C}_M\boldsymbol{\Omega}, \delta \boldsymbol{\theta}^\prime\right)_\mathcal{B} - \left( \boldsymbol{\Lambda}\boldsymbol{C}_F\boldsymbol{\Gamma}, \mathbf{S}\left(\delta \boldsymbol{\theta}\right)\mathbf{r}^\prime \right)_\mathcal{B} + \left(\mathbf{S}\left(\mathbf{w}\right)\mathbf{c}_\rho\mathbf{w}+\mathbf{c}_\rho \mathbf{\dot{w}}, \delta \boldsymbol{\theta}\right)_\mathcal{B},
\end{split}
\end{equation}
 and the linear form defaults to
 
 \begin{equation}
 \mathbf{b}^b\left(\delta\boldsymbol{\eta}^b\right):=\left(\mathbf{f}, \delta \mathbf{r}\right)_\mathcal{B} + \left[\mathbf{\tilde{f}}^T, \delta \mathbf{r}\right|_0^l + \left( \mathbf{m}, \delta \boldsymbol{\theta}\right)_\mathcal{B} + \left[\mathbf{\tilde{m}}^T\delta \boldsymbol{\theta}\right|_0^l.
 \end{equation} 
 Here, we use the external forces $\mathbf{f}$, the external moments $\mathbf{m}$, the point forces $\mathbf{\tilde{f}}$, the point moments $\mathbf{\tilde{m}}$, the angular velocity vector $\mathbf{w}$ such that $\mathbf{S}\left(\mathbf{w}\right)=\dot{\boldsymbol{\Lambda}}\boldsymbol{\Lambda}^T$, and the material curvature vector $\boldsymbol{\Omega}$ such that $\mathbf{S}\left(\boldsymbol{\Omega}\right)\mathbf{a}=\boldsymbol{\Lambda}^T\boldsymbol{\Lambda}^\prime\mathbf{a} \text{ for all } \mathbf{a}\in \mathbb{R}^3$.
 
 For the above integrals to be well defined, the appropriate function space for test and trial functions has to take the form of the product space $\mathcal{V_B}:=\mathcal{V}^\text{1}_r\times \mathcal{V}_\theta$.
 
 \subsubsection{Torsionfree Beam Model}
 \label{subsubsec:tf}
 In the case of initially straight beams with isotropic cross-section, under the assumption of vanishing shear strains, and assuming that torsional components of distributed and discrete external moments acting on the beam are negligible, the SR beam formulation can be simplified to a torsionfree (TF) model. The following formulation was originally introduced in \cite{Meier2015} for static problems, and extended to transient problem types in \cite{meier_contact}.
 
 The above assumptions lead to exactly vanishing torsion only in static problems. Nevertheless, it is suggested in \cite{meier_contact} that under the mentioned restrictions, torsional values are "very small" also in dynamic applications. Thus, the unknown vector of rotations $\boldsymbol{\psi}$ can be removed from the system so that only the unknowns $\boldsymbol{\eta}^b:=\mathbf{r}$ remain. We then define the semi-linear form
 
 \begin{equation}
 \begin{split}
 \mathbf{a}^b\left(\boldsymbol{\eta}^b; \delta\boldsymbol{\eta}^b\right) & := \left(\rho^b \mathbf{A\mathbf{\ddot{r}}}, \delta\mathbf{r}\right)_\mathcal{B} + \left(EA \dfrac{\left(\left\|\mathbf{r}^\prime\right\|-1\right)\mathbf{r}^\prime}{\left\|\mathbf{r}^\prime\right\|}, \delta\mathbf{r}^\prime\right)_\mathcal{B}\\ &+ \left(\dfrac{EI}{\left\|\mathbf{r}^\prime\right\|^4}\mathbf{S}\left(\mathbf{r}^\prime\right)\mathbf{r}^{\prime\prime}, \mathbf{S}\left(\delta \mathbf{r}^\prime\right)\mathbf{r}^{\prime\prime}+\mathbf{S}\left(\mathbf{r}^\prime\right)\delta \mathbf{r}^{\prime\prime}\right)_\mathcal{B} \\
  &- \left(\dfrac{2EI}{\left\|\mathbf{r}^\prime\right\|^6}\mathbf{S}\left(\mathbf{r}^\prime\right)\mathbf{r}^{\prime\prime}, \delta \left(\mathbf{r}^{\prime T}  \mathbf{r}\right) \mathbf{S}\left(\mathbf{r}^\prime\right)\mathbf{r}^{\prime\prime}\right)_\mathcal{B} - 
 \left(\mathbf{m},\dfrac{\mathbf{S}\left(\mathbf{r}^\prime\right)}{\left\|\mathbf{r}^\prime\right\|^2} \delta\mathbf{r}^\prime\right)_\mathcal{B} \\
 &- \left[\dfrac{\tilde{\mathbf{m}}\mathbf{S}\left(\mathbf{r}^\prime\right)\delta \mathbf{r}^\prime}{\left\|\mathbf{r}^\prime\right\|^2}\right|_0^l,
 \end{split}
 \end{equation}
 and the linear form
 
 \begin{equation}
 \begin{split}
\mathbf{b}^b\left(\delta\boldsymbol{\eta}^b\right):=\left(\mathbf{f}, \delta \mathbf{r}\right)_\mathcal{B}+\left[\mathbf{\tilde{f}}^T, \delta \mathbf{r}\right|_0^l.
 \end{split}
 \end{equation}
 The removal of rotational unknowns from the system of PDEs through enforcement of the TF constraint comes with the cost of additionally required smoothness of the centerline curves, namely $\mathbf{r}\in\mathcal{V_B}:=\mathcal{V}^\text{2}_r$.
 
 \subsection{Navier-Stokes Equations}
 \label{subsec:nse}
 
 To model the fluid field, the instationary, incompressible Navier-Stokes equations for Newtonian fluids on fixed meshes are used.
 
 In the following, $H_0^1\left(\Omega_f\right)$ represents the standard Sobolev space on the fluid domain $\Omega_f$ with zero trace on the boundary $\partial \Omega_f$, and for a domain $X$ we will frequently use the notation $\left(\cdot , \cdot \right)_{X}:=\left(\cdot , \cdot \right)_{L^2\left(X\right)}$ to denote the $L^2$ inner product on the domain $X$. The fluid boundary $\partial\Omega_f = \Gamma^f_N \cup \Gamma^f_D$ can be partitioned in a Neumann boundary $\Gamma^f_N$, on which a traction $\mathbf{h}^f$ is prescribed, and a Dirichlet boundary $\Gamma^f_D$, for which a function $\mathbf{v}^f_D$ with prescribed velocity values on $\Gamma^f_D$ is introduced. Further, let $\mathcal{V}_p = \left\{p\in L^2\left(\Omega_f\right) | \left\|p\right\|_{L^2\left(\Omega^f\right)} = 0 \right\}$ be the space of normalized pressure solutions.
 
 For the fluid field, we introduce the semi-linear form
 
  \begin{equation}
  	\begin{split}
  \mathbf{a}^f\left(\mathbf{v}^f, p^f; \delta\mathbf{v}^f, \delta p^f\right) := \rho^f \left(\dfrac{\partial \mathbf{v}^f}{\partial t}, \delta \mathbf{v}^f\right)_{\Omega_f} + 2\gamma^f \left(\boldsymbol{\mathcal{E}}\left(\mathbf{v}^f\right), \boldsymbol{\nabla} \mathbf{v}^f\right)_{\Omega_f} - \left(p^f, \boldsymbol{\nabla} \cdot \delta\mathbf{v}^f\right)_{\Omega_f} \\
  + \rho^f\left(\left(\mathbf{v}^f\cdot \boldsymbol{\nabla}\right)\mathbf{v}^f, \delta \mathbf{v}^f\right)_{\Omega_f} + \left(\boldsymbol{\nabla}\cdot \mathbf{v}^f, \delta p^f\right)_{\Omega_f},
  \end{split}
  \end{equation}
  and the linear form
  
\begin{equation}
\mathbf{b}^f\left(\delta \mathbf{v}^f\right):=\rho^f\left(\mathbf{f}^f, \delta \mathbf{v}^f\right)_{\Omega_f} + \left(\mathbf{h}^f, \delta \mathbf{v}^f\right)_{\Gamma_N^f},
\end{equation}
   with the strain rate tensor $\boldsymbol{\mathcal{E}}\left(\mathbf{v}^f\right)=\dfrac{1}{2}\left(\nabla \mathbf{v}^f+\left(\nabla\mathbf{v}^f\right)^T\right)$, a body force $\mathbf{f}^f$, the dynamic viscosity $\gamma^f$, and the fluid density $\rho^f$, respectively.
   
   In the case of a divergence-free initial velocity field $\mathbf{v}_0^f$, the behavior of the fluid on the domain $\Omega_f$ is fully described by
   
 \begin{problem}
 	\label{prob:nse}
	Find $\left(\mathbf{v}^f, p^f\right)\in L^2\left(I, H_0^1\left(\Omega_f\right)^3+\mathbf{v}_D\right)
 	\times L^2\left(I, \mathcal{V}_p \right)$, with $\mathbf{v}^f=\mathbf{v}_0$ a. e. for $t=0$, such that
 \begin{equation}
   	\int\limits_0^T \mathbf{a}^f\left(\mathbf{v}^f, p^f; \delta\mathbf{v}^f, \delta p^f\right)
 - \mathbf{b}^f\left(\delta \mathbf{v}^f\right) \text{ d}t=0 \quad \forall \quad \left(\delta \mathbf{v}^f, \delta p^f\right) \in H_0^1\left(\Omega_f\right)^3\times \mathcal{V}_p.
   \end{equation}
 \end{problem}
   
   \subsection{Mixed-Dimensional Fluid-Beam Interaction}
   \label{subsec:fbi}
   
   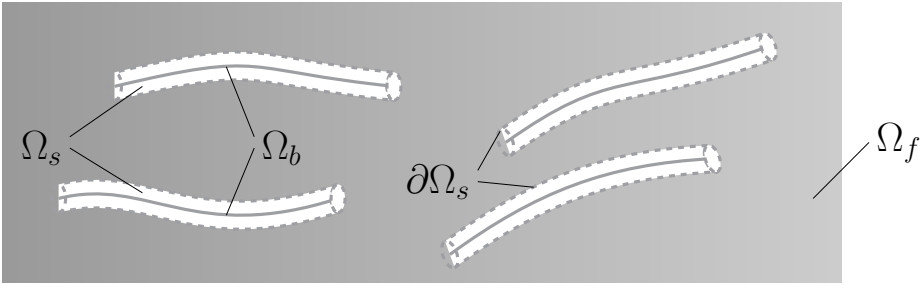
\begin{figure}[h]
   	\begin{center}
   	\resizebox{0.8\textwidth}{!}{
   \begin{tikzpicture}
   \fill[left color=black!40, right color=black!20] (0,0.5) rectangle (15,5.5); 
   \pgfmathsetseed{2015}
   \foreach \x in {1,...,2}
   \draw (\x, 2*\x) pic[rotate=-90] {bent cylinder};
      \foreach \x in {1,...,2}
      \draw (7 + \x, -1 + 2*\x) pic[rotate=-70] {bent cylinder};    
   \draw (14.5, 2) -- (15.5, 3) node[anchor = west] {\huge{$\Omega_f$}}; 
   \draw (2.5, 2.1) -- (1.2, 2.9) node[anchor = east] {\huge{$\Omega_s$}};
   \draw (2.5, 4) -- (1.2, 3.1);
   \draw (4, 1.7) -- (4.5, 2.9) node[anchor = west] {\huge{$\Omega_b$}};
   \draw (4, 4.35) -- (4.5, 3.1);
   \draw (9.5, 2.2) -- (8.5, 2.3) node[anchor = east] {\huge{$\partial\Omega_{s}$}};
   \draw (8.9, 3.2) -- (8.5, 2.5);
   \end{tikzpicture}
    }
   \caption{Domain of a fluid-beam interaction problem}
   \end{center}
   \end{figure}
   
   The single field equations introduced within the last subsections can now be used to formulate the coupled fluid-beam interaction problem. Generally, the multi-physics nature of FSI problems leads to additional challenges compared to single field equations. Among these challenges is the fact, that the single field equations are based on different primal variables, i.e. the velocity field $\mathbf{v}^f$ for the Navier-Stokes equations, and the displacement field or, more accurately in our case, the beam position $\mathbf{r}$ for the structure. For IBM type methods it is thus customary to introduce the structure velocity $\mathbf{v}^s$ in order to formulate the kinematic coupling condition
   
   \begin{equation}
   \label{eq:vel_coupling_solid}
   \mathbf{v}^f=\mathbf{v}^s \qquad \text{a. e. on } \partial\Omega_{s},
   \end{equation}
 to ensure continuity of the velocity on the fluid-structure interface $\partial\Omega_{s}$.
   
   Here, $\Omega_s$ represents the fictitious domain, which would be occupied by a fully resolved beam. The aforementioned coupling equation leads to a surface-to-volume coupling approach, as adopted in \cite{Henshaw2015, huang2019}, thus triggering a growing complexity in the computations of the coupling conditions.
      
      Based on the results in \cite{Baaijens2001} and under the assumption of sufficiently small beam radii, the following simplifications can be argued to be valid:
      \begin{enumerate}
      	\item The coupling of the two fields can be realized on the beam centerline.
      	\item Rotational effects on the fluid flow are negligible.
      \end{enumerate}

      For the beam velocity $\mathbf{v}^b:= \dot{\mathbf{r}}$, \eqref{eq:vel_coupling_solid} can be rewritten as

      \begin{equation}
      \label{eq:vel_coupling}
      \mathbf{v}^f= \mathbf{v}^b \circ \mathbf{r}^{-1} \qquad \text{a. e. on } \Omega_b,
      \end{equation}
      where $\mathbf{r}^{-1}$ serves as projection of the beam parameter space $\left[0,l\right]$ onto the beam's current geometry $\Omega_b=\Omega_b\left(t\right):=\mathbf{r}\left(t, \left[0,l\right]\right)$ in three-dimensional space. Equivalently, this condition can be written as
                
      \begin{equation}
      \label{eq:vel_coupling_reference}
      \mathbf{v}^f\circ \mathbf{r}= \mathbf{v}^b \qquad \text{a. e. on } \left[0,l\right],
      \end{equation}
      on the parameter space $\left[0,l\right]$. 
     
     Strong enforcement of the kinematic condition may lead to spurious effects on the fluid pressure field for immersed boundary methods because of its impact on the divergence condition. We will thus impose this condition in a weak sense. A common weak constraint enforcement technique is the Lagrange multiplier method as e.g. used in \cite{Kloppel2012, Baaijens2001, mayr2014}. Application of Lagrange multipliers generally leads to a saddle point problem, which has to adhere to the inf-sup condition \cite{brezzi1990} for a stable solution. In general, for embedded finite element methods the form of such a stable Lagrange multiplier space depends on the position of the embedded mesh relative to the background mesh \cite{bechet2009, hautefeuille2012}. For the case of mixed-dimensional problems, the form of such stable Lagrange multipliers is not yet well-studied. Alternatively, analogous to Nitsche's method for classical equal-dimensional embedded finite element problems \cite{sanders2012, hautefeuille2012}, stabilized Lagrange multiplier methods can also be applied to mixed-dimensional embedded finite element problems \cite{kuchta2020, kerfriden2020}.
     
     Nevertheless, in order to analyze the impact of this novel approach of coupling the 3D Navier-Stokes equations to 1D beam theory directly on the centerline without the need to address the additional challenge of constructing a stable Lagrange multiplier space or introducing additional stabilization terms, we refrain from using Lagrange multipliers here. Instead, we want to impose the kinematic coupling condition \eqref{eq:vel_coupling} weakly via the penalty method with a penalty parameter $\epsilon$. Constraint enforcement via the penalty method generally leads to a more or less moderate violation of the kinematic condition. On the other hand, the method comes with the advantage of simplicity of the implementation and, more importantly, robustness with regard to stability, independently of the position of the beam. Further development to include stabilized Lagrange multipliers into the proposed method is left for future work, should challenges in connection with the penalty method, such as well-known deterioration of the condition number of the system matrix, arise.
      
      Under the above mentioned assumption of sufficiently small beam radii, the dynamic coupling condition ensuring continuity of tractions takes the form
      
      \begin{equation}
      \label{eq:force_equilibrium}
      \left(\mathbf{f}, \delta\mathbf{r}\right) + \lim\limits_{\partial\Omega_{s}\rightarrow \Omega_b} \left(\boldsymbol{\sigma}\left(\mathbf{v}^f\right)\cdot \mathbf{n}, \delta \mathbf{v}^f\right)_{\partial\Omega_{s}} = 0.
      \end{equation}
      
      It is not obvious how the result of the second term looks in the limit case. However, we postulate, using the restriction operator $\boldsymbol{\Pi}:H^1\left(\Omega_f\right)^3 \rightarrow L^2\left(\Omega_b\right)$, the second term in \eqref{eq:force_equilibrium} in the limit case leads to a one-dimensional integral 
       
       \begin{equation}
       \int_{\Omega_b}\mathbf{f}_{FBI}^f\left(\mathbf{v^f}\right)\cdot \boldsymbol{\Pi}\delta\mathbf{v}^f\text{ d}s,
       \end{equation}
      where the result of the operator $\mathbf{f}_{FBI}^f\left(\mathbf{v^f}\right):H^1\left(\Omega_f\right)^3 \rightarrow L^2\left(\Omega_b\right)$ represents a line force acting on the fluid. For the treatment via the penalty method, postulating the existence of such a line force suffices. Nevertheless, for coupling via a range of regularized Lagrange multiplier methods such as Nitsche type methods, further analysis of the exact form of the resulting line force is necessary. This question is closely related to the existence of a sufficiently smooth restriction operator $\boldsymbol{\Pi}$.
      
      \begin{remark}
      Here, $\boldsymbol{\Pi}$ is necessary, since the integral over the one-dimensional curve $\Omega_b$ is obviously not well-defined for $\delta\mathbf{v}^f\in H_0^1\left(\Omega_f\right)^3$. In contrast to well-known trace theorems such as \cite{jerison1981}, which postulate existence of such a trace operator on smooth boundaries of codimension one, existence conditions on the restriction operator $\boldsymbol{\Pi}$ in the context of a greater dimensionality gap are not yet well-studied \cite{kuchta2019}. As one of the first publications addressing the lack of trace-type theorems for mixed-dimensional problems with codimension two, Kuchta et al. \cite{kuchta2020} show sufficient regularity of such a restriction operator in the context of a mixed-dimensional model problem via averaging over a three-dimensional domain around the embedded manifold.
            
      Even though a theoretical analysis of the well-posedness of the continuous problem in weak form would certainly represent a firm basis for further work, we will refrain here from any theoretical existence or regularity analysis. Instead, we will validate the choice of the coupling domain and illustrate its challenges by selected numerical examples in Section \ref{sec:examples}.
      \end{remark}
            
    Finally, we introduce constraint \eqref{eq:vel_coupling_reference} with the penalty method, as shown in \cite{penalty}, using a penalty contribution that increases linearly with respect to the constraint violation, by building the derivatives of an abstract quadratic penalty functional:
    
    \begin{equation}
    \dfrac{\epsilon}{2} \int\limits_0^l \left(\boldsymbol{\Pi v}^f\circ \mathbf{r}-\mathbf{v}^b\right) \cdot \left(\boldsymbol{\Pi v}^f\circ \mathbf{r}-\mathbf{v}^b\right) \text{ d}s.
    \end{equation}

    Here, the slope of the linear penalization term depends on the value of the penalty parameter $\epsilon$. This leads to the full monolithic coupling problem
            
      \begin{problem}
      	\label{prob:fbi}
      	Find $\left(\mathbf{v}^f, p^f, \boldsymbol{\eta}^b\right)\in L^2\left(I, H^1\left(\Omega_f\right)^3+\mathbf{v}_D\right)\times L^2\left(I, \mathcal{V}_p \right)\times L^2\left(I, \mathcal{V_B}\right)$, with $\mathbf{v}^f=\mathbf{v}_0$, $\boldsymbol{\eta}^b=\boldsymbol{\eta}_0$ a. e. for $t=0$, and $\left\|\mathbf{r}^\prime\left(0, \cdot\right)\right\|_{\mathbb{R}^3}=1$ a. e. on $\left[0,l\right]$, such that
      	\begin{equation}
      	\label{eq:fbi}
      	\begin{split}
      		\int\limits_0^T \mathbf{a}^f\left(\mathbf{v}^f, p^f; \delta\mathbf{v}^f, \delta p^f\right)
      		- \mathbf{b}^f\left(\delta \mathbf{v}^f\right)  + \epsilon \int\limits_0^l \left(\boldsymbol{\Pi v}^f\circ \mathbf{r}-\mathbf{v}^b\right) \cdot \boldsymbol{\Pi}\delta\mathbf{v}^f\circ \mathbf{r} \text{ d}s \text{ d}t=0, \\
      		\int\limits_0^T \mathbf{a}^b\left(\boldsymbol{\eta}^b; \delta\boldsymbol{\eta}^b\right) - \mathbf{b}^b\left(\delta\boldsymbol{\eta}^b\right) - \epsilon \int\limits_0^l \left(\boldsymbol{\Pi v}^f\circ \mathbf{r}-\mathbf{v}^b\right) \cdot\delta\mathbf{r} \text{ d}s \text{ d}t=0,
      	\end{split}
      	\end{equation}
      	for all $\left(\delta\mathbf{v}^f, \delta p^f, \delta \boldsymbol{\eta}^b\right)\in H_0^1\left(\Omega_f\right)^3\times \mathcal{V}_p\times \mathcal{V_B}$.
      \end{problem}
   
\section{Discretization and Numerical Integration}
\label{sec:numerics}

The PDEs of both fields introduced in Section \ref{sec:equations} are spatially discretized using the FEM, and the evolution in time is modeled using a finite difference time stepping scheme. We want to briefly recount particular challenges arising for the respective fields, and point out how they were overcome within the present work. Afterwards, the numerical treatment of the coupling terms will be presented in more detail.

\subsection{Fluid Field}
Numerical treatment of the incompressible Navier-Stokes equations presented in Section \ref{subsec:nse} with the FEM leads to a mixed formulation, which generally has to adhere to the inf-sup stability condition, cf. \cite{brezzi1990}. Nevertheless, we will take the alternative approach and circumvent the inf-sup condition by stabilizing linear equal-order hexahedral finite elements using the PSPG/SUPG stabilization method with an additional div-grad stabilization term \cite{hansbo1990, franca1992}. Here, only the PSPG method actually circumvents the inf-sup condition. The SUPG method prevents numerical instabilities in convection-dominated flows, while the grad-div term improves well-posedness as well as solution accuracy of the Navier-Stokes equations. A detailed description and recounting of all stabiliziation terms within the used implementation can be found in \cite{schott2015}.

Thus the unknown fields of Problem \ref{prob:nse} are discretized as

\begin{eqnarray}
\label{eq:ansatz_fluid}
\mathbf{v}^f_h=\sum_{k=1}^{n^f}N_k\mathbf{\hat{v}}^{f,k}_h \qquad \text{and} \qquad p^f_h=\sum_{k=1}^{n^f}N_k{\hat{p}}^{f,k}_h,
\end{eqnarray}
where $n^f$ is the number of fluid nodes, the $N_k$ represent the linear shape functions, and $\mathbf{\hat{v}}^{f,k}_h$ and ${\hat{p}}^{f,k}_h$ denote the nodal degrees of freedom of the velocity and pressure at node $k$, respectively.

To discretize the Navier-Stokes equations in time, a one-step-$\theta$ time stepping scheme is employed for the examples in Section \ref{sec:examples}.

\subsection{Beams}
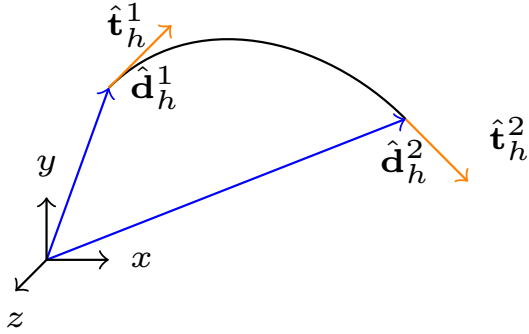
\begin{figure}[!h]
    \centering
    \resizebox{0.5\textwidth}{!}{\begin{tikzpicture}

\draw (-1.6, 0) to [out=45, in=135] (0.3, -0.2);

\draw[->, orange] (-1.6, 0) -- (-1.2, 0.4) node[anchor=east]{\color{black}\tiny{$\hat{\mathbf{t}}_h^1$}};
\draw[->, orange] (0.3, -0.2) -- (0.7, -0.6) node[anchor=south west]{\color{black}\tiny{$\hat{\mathbf{t}}_h^2$}}; 

\draw[->, blue] (-2, -1.1) -- (-1.6, 0) node[anchor=west]{\color{black}\tiny{$\hat{\mathbf{d}}_h^1$}};
\draw[->, blue] (-2, -1.1) -- (0.3, -0.2) node[anchor=north]{\color{black}\tiny{$\hat{\mathbf{d}}_h^2$}};

\draw[->] (-2, -1.1) -- (-1.6, -1.1) node[anchor=west]{\tiny{$x$}};
\draw[->] (-2, -1.1) -- (-2, -0.7) node[anchor=south]{\tiny{$y$}};
\draw[->] (-2, -1.1) -- (-2.2, -1.3) node[anchor=north]{\tiny{$z$}};

\end{tikzpicture}}
    \caption{Visualization of a Hermite beam centerline interpolation}
    \label{fig:beam_element}
\end{figure}

The additionally required smoothness of the centerline representation within the torsionfree beam formulation leads to the necessity of $C^1$-continuous shape functions. This is realized by using 3rd-order Hermite shape functions for the beam centerline, as described in \cite{Meier2014}. Even though these $C^1$-continuous shape functions are only necessary for the torsionfree beam element, we will also apply them to the centerline interpolation of the SR finite element used in the numerical examples in Section \ref{sec:examples}. As shown in Figure \ref{fig:beam_element}, this leads to a two-noded element with six centerline degrees of freedom (DoFs) per node and the shape functions $H^d_k$ for the nodal positions $\hat{\mathbf{d}}_h$ in all three dimensions, and $H^t_k$ for the nodal tangents $\hat{\mathbf{t}}_h$ in all three dimensions.

The description of the discrete beam position thus takes the form

\begin{equation}
\label{eq:ansatz_beam}
\mathbf{r}_h=\sum_{k=1}^{n^b}H^d_k\hat{\mathbf{d}}_h + \dfrac{l_{ele}}{2}\sum_{k=1}^{n^b}H^t_k\hat{\mathbf{t}}_h,
\end{equation}
with the number of beam-centerline nodes $n^b$ and the initial beam element length $l_{ele}$.

As argued in Section \ref{subsec:fbi}, for relatively slender beams, it is valid to neglect the effect of beam rotations on the fluid flow. Thus, only the centerline terms, which are independent of the employed beam theory, appear in the coupling equations. Therefore, we skip the details of the cross-section rotation discretization here. The interested reader may be referred to \cite{meier_contact} for details on the applied Petrov-Galerkin method for discretization in space, and to \cite{bruels2010, bruels2012} for the Lie-group Generalized-$\alpha$ time stepping scheme based on multiplicative updates, which is applied to the rotational degrees of freedom of the employed beam finite elements.

The time evolution of the beam centerline position for all examples in Section \ref{sec:examples}, including SR as well as torsionfree beam elements, is discretized using a Generalized-$\alpha$ time integration scheme \cite{genalpha} with the spectral radius set to $\rho_\infty = 1$, and all parameters set accordingly to produce a second-order accurate time stepping scheme.

\subsection{Numerical Treatment of the Coupling Condition}

After discretization of both, the three-dimensional incompressible isothermal Navier-Stokes equations and the beam formulation, two distinct non-matching meshes emerge. The fixed fluid mesh takes on the role of a background mesh. The beam mesh, oriented along the beams' centerlines, is superimposed, leading to the use of an embedded mesh approach. The proposed coupling discretization can be characterized as a Gauss-point-to-segment approach, as e.g. commonly used in contact mechanics. Here, the constraint equation \eqref{eq:vel_coupling_reference} is evaluated at each quadrature point of the beam. Future work will focus on the extension to a segment-to-segment (STS) based approach. In our previous work \cite{steinbrecher2020}, it was shown that realizations of the mesh tying of a beam to a solid volume can benefit from such a STS based approach in terms of well-posedness and the avoidance of locking effects. Nevertheless, the numerical examples in Section \ref{sec:examples} will show that the used GPTS approach is more than sufficient for a multitude of application problems.

In the following, we will summarize the beam element shape functions at node $k$, including positional shape functions $H_k^d$ and tangential shape functions $H_k^t$, as $H_k$. Inserting \eqref{eq:ansatz_beam} and \eqref{eq:ansatz_fluid} into the coupling contributions introduced in Problem \ref{prob:fbi} gives rise to four blocks of coupling contributions to the discretized system:

\begin{equation}
\begin{split}
&\epsilon\mathbf{K}_\mathcal{BB}\left(k, j\right):=\epsilon\int\limits_0^l H_kH_j\text{ d}s, \qquad \qquad \epsilon\mathbf{K}_\mathcal{FF}\left(k, j\right):=\epsilon\int\limits_0^l\left(N_k\circ\chi\right) \left(N_j\circ\chi\right) \text{ d}s, \\
&\epsilon\mathbf{K}_\mathcal{BF}\left(k, j\right):=\epsilon\mathbf{K}_\mathcal{FB}\left(j,k\right):=\epsilon\int\limits_0^lH_k\left(N_j\circ\chi\right)\text{ d}s,
\end{split}
\label{eq:discretization}
\end{equation}

with $\chi$ denoting the projection of a point in the parameter space of the beam centerline to the corresponding point in the parameter space of the discretized fluid volume. Subscripts $\mathcal{BB}$ and $\mathcal{FF}$ denote beam and fluid contributions, respectively. Mixed subscripts $\mathcal{FB}$ and $\mathcal{BF}$ refer to the mixed coupling interaction contributions.
Note that here, the projection $\chi$ is achieved by computing the current position $\mathbf{r}_h$ at each Gauss point of the discretized beam centerline curve. Thus, the projection $\chi$ introduces a hidden dependency on the current deformation of the beam, and adds an additional nonlinearity to the system. For the discretized monolithic formulation introduced in Problem \ref{prob:fbi}, the addition of the above coupling matrices to the system matrix is straight forward. However, we have chosen to use a partitioned FSI solution scheme. The exact way the coupling matrices in \eqref{eq:discretization} are added to the numerical system, as well as details on the algorithmic treatment of the resulting nonlinearity will be given in Section \ref{sec:algo}.

\begin{figure}[!h]
	\begin{center}
		\includegraphics[width=0.8\textwidth]{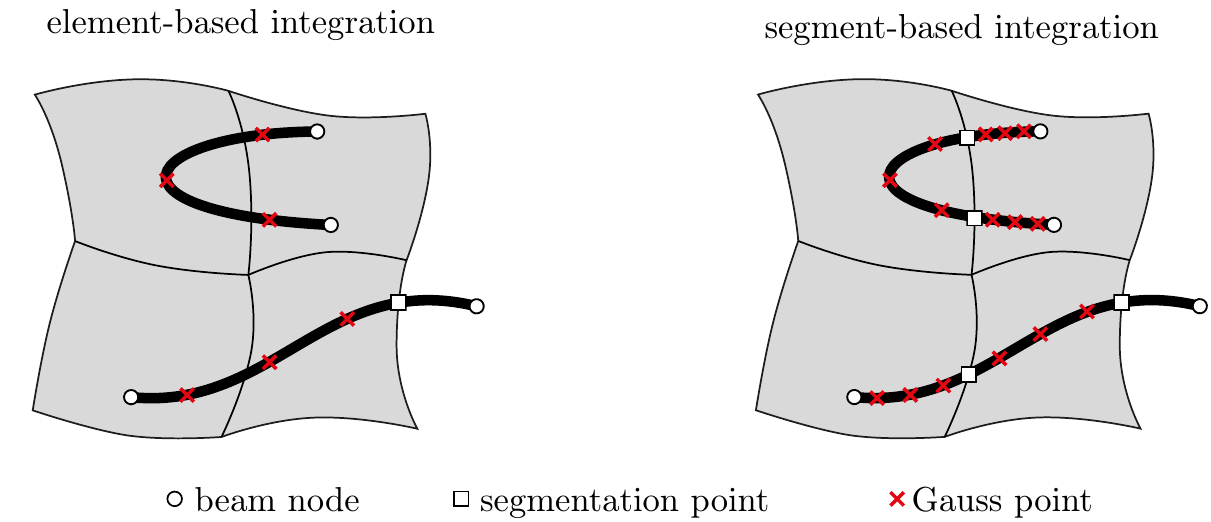}
	\end{center}
	\caption{Segmentation of the beam element for stable and accurate numerical integration as presented in \cite{steinbrecher2020}: circles and crosses denote beam nodes and quadrature points, respectively. The squares subdivide the beam into integration segments, such that an integration cell on the beam does not cross element boundaries of the fluid mesh to not integrate across weak discontinuities.}
	\label{fig:segmentation}
\end{figure}

In order to deal with weak discontinuities at the fluid element boundaries as they arise during element-wise numerical integration of the integrals in \eqref{eq:discretization}, a more accurate segment-based integration approach to evaluate \eqref{eq:discretization} is adopted from our previous work \cite{steinbrecher2020} and visualized in Figure \ref{fig:segmentation}.

As the present problem is transient, the time discretization for the single fields is also applied to the constraint. Aiming at a temporally consistent exchange of coupling information \cite{mayr2014, joosten2010}, this leads to a linear combination of the coupling contributions $\mathbf{K}_{\circ\circ}^n$,  evaluated at the old time step $n$, and $\mathbf{K}_{\circ\circ}^{n+1}$, at the new time step $n+1$, depending on the time integration parameters for the employed fluid and structure time integration schemes, respectively. Further information on the algorithmic details will be given in the upcoming section.

\begin{remark}
\label{remark:element_size}
  Note that the assumption of relatively small beam radii also extends to the fluid element sizes. The choice of enforcing the fluid-beam coupling on the beam centerline introduces a singularity to \eqref{eq:fbi}. The fluid elements need to be large enough to not be able to resolve this singularity, effectively "smearing" the interface over the volume of the element as is common in classical immersed methods. The exact domain of applicability of the method with respect to the quotient of beam diameter and fluid element size also depends on the properties of the simulated fluid. The influence of the background mesh resolution on the fluid velocity resolution will be studied in section \ref{sec:examples}.
\end{remark}

\section{One-way Coupling Algorithms}
\label{sec:algo}

   \begin{figure}[h!]
    \begin{center}
        \resizebox{\textwidth}{!}{
            \begin{tikzpicture}
            \node[cloud] (fluid) {Solve fluid problem $\mathbf{f}_h^{\mathcal{F}, n+1, i + 1}:=\mathcal{F}\left(\mathbf{v}_b^{n+1, i}\right)$};
            \node[cloud, node distance=12cm, left of=fluid] (beam) {Solve beam problem $\mathbf{v}_b^{n+1, i}:=\mathcal{B}\left(\mathbf{f}_h^{\mathcal{B}, n+1, i}\right)$};
            \node[below of=beam, node distance=2cm] {fluid-to-beam coupling};
            \node[above of=fluid, node distance=2cm] {beam-to-fluid coupling};
            
            \path[line] (fluid) to [out=-150, in=-30] node[above] {Increment $i$} (beam);
            \path[line] (fluid) to [out=-150, in=-30] node[below] {Calculate reaction force $\mathbf{f}_h^{\mathcal{B}, n+1, i}$} (beam);
            \node at (-6.2,-3.1) {acting on the beam};
            \path[line] (beam)  to  [out=30, in=150] node[above] {$\mathbf{r}^{n+1, i}_h, \mathbf{v}_h^{b, n+1, i}$} (fluid);
            
            \path[line] (fluid) to [out=-130, in=-50, looseness = 2] node[below] {$j^{\text{th}}$ Newton iteration} (fluid);
            \path[line] (beam) to [out= 130, in=50, looseness = 2] node[above] {$k^{\text{th}}$ Newton iteration} (beam);
            \end{tikzpicture}
        }
        \caption{Visualization of the one-way fluid and solid coupling schemes. In the visualization, $\mathcal{B}$, $\mathcal{F}$ represent suitable operators for the solution of the nonlinear beam problem, and for the solution of the nonlinear fluid problem, respectively.}
        \label{fig:fsialgo}
    \end{center}
\end{figure}
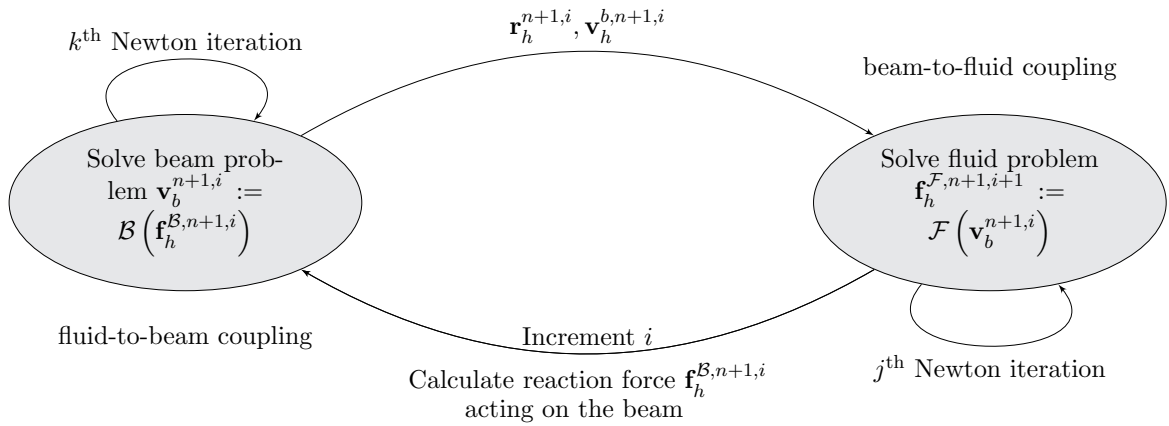

The solution algorithm is set up as a partitioned FSI algorithm with the counter $i$ representing the coupling iteration. The long-term goal of our current work is the application of the proposed mixed-dimensional coupling approach to a fully coupled fluid-beam interaction problem. Nevertheless, in order to analyze the implications and limitations stemming from the numerical embedding of a reduced-dimensional structure in three-dimensional flow, here, we first want to introduce two one-way coupled algorithmic variants. On the one hand, this analysis serves as a validation of the different components, which can then be combined to a fully two-way coupled fluid-beam interaction algorithm. On the other hand, as will be shown in Section \ref{sec:examples}, there exist numerous application scenarios, in which the one-way coupling variants proposed here represent valid models at rather low computational complexity. This highlights their value not only as a work-in-progress, but particularly as stand-alone models for a broad spectrum of practically relevant examples.

In the upcoming subsections, we will explain the implications for the two different one-way coupling schemes depicted in Figure \ref{fig:fsialgo}. Nevertheless, some details are valid for both variants. Firstly, the single fields are solved using a Newton-Raphson algorithm. The number of Netwon iterations within the single fields will be represented by the counter $j$ and $k$ for the fluid and structure field, respectively. Secondly, to capture the exact FBI boundary, i.e. the position of the beam centerline, an octree-based search is performed after each change in the state of the solution of the nonlinear beam problem. This way, all fluid element nodes in a prescribed vicinity of the displaced beam element nodes are found. The segmentation procedure visualized in Figure \ref{fig:segmentation} is applied to all resulting pairs of intersecting beam and fluid elements. If a valid segment is found within such a pair, its contributions are added to the respective matrices in \eqref{eq:discretization}. Otherwise, the pair does not contribute to the coupling terms.

In order to not complicate the notation unnecessarily, we will drop the iteration counter indices $i$, for the partitioned algorithm, and $j$ as well as $k$, for the field-specific Newton-Raphson solver, to indicate independence of a variable with respect to that index.

\subsection{One-Way Coupling for Rigid Beams}
\label{subsec:solid-to-fluid}

First, we analyze the special case of immersed rigid beams. Assuming that the effect of the fluid flow on the beam is negligible allows to use a one-way coupling algorithm which only affects the fluid partition. We will denote this one-way coupling scheme as beam-to-fluid coupling variant within the remaining part of the paper.

Motivated by classical Dirichlet-Neumann partitioned schemes, where $\Omega_f$ acts as Dirichlet and $\Omega_b$ as Neumann Partition, the discrete penalty constraint at time step $n+1$
\begin{equation}
\label{eq:weakdirichletconstraint}
\epsilon\mathbf{K}^{n+1}_\mathcal{FF}\mathbf{v}^{f, n+1}_h=\epsilon\mathbf{K}^{n+1}_\mathcal{FB}\mathbf{v}^{b, n+1}_h,
\end{equation}
is enforced on the fluid partition for a prescribed discrete beam velocity $\mathbf{v}^{b, n+1}_h$. Here, the coupling variable $\mathbf{v}^{b, n+1}_h$ is computed by the beam's time integration scheme. Constraint \eqref{eq:weakdirichletconstraint} can be interpreted as the weak enforcement of the Dirichlet constraint

\begin{equation}
	\mathbf{v}^{f, n+1}_h=\bar{\mathbf{v}}^{b, n+1}_h:=\left(\mathbf{K}_\mathcal{FF}^{n+1}\right)^{-1}\mathbf{K}^{n+1}_\mathcal{FB}\mathbf{v}^{b, n+1}_h,
\end{equation}
where $\bar{\mathbf{v}}^{b, n+1}_h$ can be interpreted as the projection of the beam velocity onto the fluid mesh. Note that the matrices $\mathbf{K}^{n+1}_\mathcal{FF}$ and $\mathbf{K}^{n+1}_\mathcal{FB}$ in \eqref{eq:weakdirichletconstraint} only depend on the geometry of the overall system. Because the fluid is solved in a purely Eulerian frame, the above mentioned coupling matrices are constant within the nonlinear solution step of the fluid partition. Accounting for the time discretization, the following equation is added to the fluid system in iteration step $j+1$ of the Newton-Raphson method:

\begin{equation}
\begin{split}
\theta\epsilon\mathbf{K}^{n+1}_\mathcal{FF}\mathbf{v}^{f, n+1, j+1}_h &= \theta\left(\epsilon\mathbf{K}^{n+1}_\mathcal{FB}\mathbf{v}^{b, n+1}_h - \epsilon\mathbf{K}^{n+1}_\mathcal{FF}\mathbf{v}^{f, n+1, j}_h\right) \\
&+ \left(1-\theta\right)\left(\epsilon\mathbf{K}^{n}_\mathcal{FB}\mathbf{v}^{b, n}_h - \epsilon\mathbf{K}^{n}_\mathcal{FF}\mathbf{v}^{f, n}_h\right).
\end{split}
\end{equation}

At the same time, the fluid solution has no effect on the structure within this one-way coupling variant. The partitioned algorithm will thus converge after only one iteration.

\subsection{One-Way Coupling for Light Fibers}
\label{subsec:fluid-to-solid}
Conversely, we analyze the special case of freely moving, light fibers, for which the effect on the fluid flow can be neglected. Within the remainder of this paper, we will refer to this one-way coupling scheme as the fluid-to-beam coupling variant.

Once more, motivated by classical Dirichlet-Neumann partitioned schemes, where the fluid domain is treated as Dirichlet partition, while the structure domain takes on the role of the Neumann partition, the coupling condition \eqref{eq:vel_coupling_reference} is now applied to the beam in the form of the discrete interaction force

\begin{equation}
\label{eq:beam_force}
	\mathbf{f}_h^{\mathcal{B}, n+1, i+1}:=\epsilon\mathbf{K}_\mathcal{BB}^{n+1, i+1}\mathbf{v}_h^{b, n+1, i+1} - \epsilon\mathbf{K}_\mathcal{BF}^{n+1, i+1}\mathbf{v}_h^{f, n+1},
\end{equation}
within every iteration $i+1$ of the partitioned algorithm.
For the employed generalized-$\alpha$ time integration scheme, this, in turn, leads to the contribution

\begin{equation}
    \label{eq:genalpha}
    \left(1-\alpha^f\right)\mathbf{f}_h^{\mathcal{B}, n+1, i+1} + \alpha^f\mathbf{f}_h^{\mathcal{B}, n},
\end{equation}
to the residual of the structure system. Due to the partitioning, the force is constant during the solution of the structure problem, which leads to a neglect of the change in geometry of the FSI interface. This nonlinearity is thus not treated within the Newton solver for the beam system, but successively updated through iterations between the two fields. The handling of the kinematic coupling constraint as a Neumann condition within a fixed-point iteration enables a flexible choice of the beam's time integration scheme as compared to a full Newton method, in which the velocity-displacement relationship would enter the linearization of the problem. Nevertheless, for very slender structures, among others, generally convergence issues of the partitioned scheme associated with the well-known added mass effect may arise. More detailed characterizations of the added-mass effect, and a comprehensive study on methods to overcome the arising challenges, can be found in \cite{Bukac2016, CAUSIN2005}. In our case, the convergence behavior of the coupling algorithm is improved via the Aitken relaxation introduced in \cite{aitken}. In typical FBI examples, due to the beam's slender geometry and, thus, its susceptibility to small forces, the beam solution is more sensitive to changes in the geometric configuration than the fluid solution. In contrast to the method outlined in \cite{kuttler2008fixed} for typical applications of classical surface-coupled FSI algorithms, we will, thus, relax the force applied to the beam instead of the  solid displacement increment and velocity given to the fluid partition.

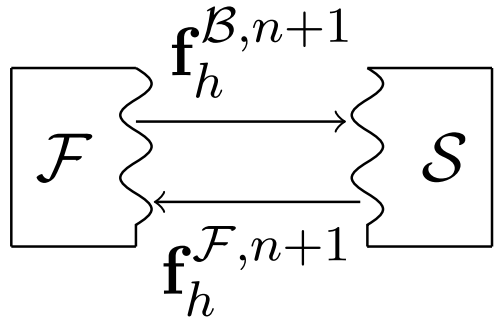
\begin{figure}[h!]
    \begin{center}
        \resizebox{0.45\textwidth}{!}{
            \begin{tikzpicture}
            \draw (-0.2,-0.5)  -- (-0.2, 0.5) node[pos = 0.5, anchor = west]{$\mathcal{F}$};
            \draw (-0.2, 0.5)  -- (0.5, 0.5);
            \draw (-0.2, -0.5)  -- (0.5, -0.5);
            \draw[snake=coil, segment aspect = 0] (0.5, 0.5)  -- (0.5, -0.5);
            \draw[ snake=coil, segment aspect = 0] (1.8, 0.5)  -- (1.8, -0.5);
            \draw (1.8, -0.5)  -- (1.8, -0.5);
            \draw (1.8, -0.5)  -- (2.5, -0.5);
            \draw (1.8, 0.5)  -- (2.5, 0.5);
            \draw (2.5, -0.5)  -- (2.5, 0.5) node[pos = 0.5, anchor = east] {$\mathcal{S}$};
            \draw[->] (0.5, 0.2) -- (1.68, 0.2) node[pos = 0.6, anchor = south] {$\mathbf{f}_h^{\mathcal{B}, n+1}$};
            \draw[->] (1.76, -0.25) -- (0.6, -0.25) node[pos = 0.5, anchor = north] {$\mathbf{f}^{\mathcal{F}, n+1}_h$};
            \end{tikzpicture}
        }
        \caption{Depiction of the FSI interaction forces between the fluid field $\mathcal F$ and the structure field $\mathcal{S}$}
        \label{fig:fbi-force}
    \end{center}
\end{figure}

\begin{remark}
	\label{remark:DN}
    
We want to give a few remarks on the connection between classical Dirichlet-Neumann algorithms and the algorithm proposed here. Transformation of \eqref{eq:fbi} leads to
\begin{equation}
    \epsilon \int\limits_0^L \left(\boldsymbol{\Pi v}^f\circ \mathbf{r} - \mathbf{v}^b\right) \cdot \boldsymbol{\Pi}\delta\mathbf{v}^f\circ \mathbf{r} \text{ d}s =\mathbf{b}^f\left(\delta \mathbf{v}^f\right) - \mathbf{a}^f\left(\mathbf{v}^f, p^f; \delta\mathbf{v}^f, \delta p^f\right),
\end{equation}
meaning that, for the forces to be in equilibrium within each time step,
\begin{equation} \mathbf{f}_h^{\mathcal{F}, n+1}:=\epsilon\mathbf{K}^{n+1}_\mathcal{FF}\mathbf{v}^{f, n+1}_h - \epsilon\mathbf{K}^{n+1}_\mathcal{FB}\mathbf{v}^{b, n+1}_h
\end{equation}
 will take on the additional force applied to the fluid because of the FBI constraint.

As depicted in Figure \ref{fig:fbi-force}, in the present case of non-matching meshes, $\mathbf{f}_h^{\mathcal{B}, n+1}$ represents the reaction force to $\mathbf{f}_h^{\mathcal{F}, n+1}$  acting on the beam mesh. Thus, the load and motion transfer scheme introduced above for one-way coupling schemes leads to a weak Dirichlet-Neumann partitioned algorithm in the case of two-way coupling. Here, the Dirichlet conditions are weakly imposed on the fluid partition while the beam domain takes on the role of the Neumann partition.

Nevertheless, this interpretation of $\mathbf{f}_h^{\mathcal{B}, n+1}$ as the FBI interaction force acting on the beam weakens in the case of one-way coupling when $\mathbf{f}_h^{\mathcal{F}, n+1}$ is assumed to be negligible. In this case, \eqref{eq:beam_force} should be interpreted as weak enforcement of the Dirichlet constraint \eqref{eq:vel_coupling_reference}, for which the nonlinearity introduced by the change in geometry is treated by the Aitken relaxation instead of a Newton method. Both interpretations will be further addressed in Section \ref{sec:examples}.
\end{remark}

\begin{remark}
Even if the assumptions of a perfectly rigid beam might not hold for a given physical setup, the numerical model of the beam-to-fluid coupling variant introduced in Section \ref{subsec:solid-to-fluid} still represents a physically meaningful and solvable situation. This is not necessarily the case for the fluid-to-beam coupling variant introduced in Section \ref{subsec:fluid-to-solid}: As argued in Remark \ref{remark:DN}, the penalty force does not represent the FSI force in the fluid-to-beam coupling case. Instead, it is simply the scaled negative constraint violation. 
Since the beam is invisible to the fluid field, violation of the assumption of a perfectly soft, freely movable, light fiber may lead not only to unphysically large forces acting on the beam but also to a deterioration of convergence behavior of the given algorithm for large penalty parameters due to the ill-posedness of the underlying continuous problem. It is thus paramount to carefully check the modeling assumptions of this one-way coupling variant before considering applying the algorithm. The applicability of the proposed approach under no or even moderate violation of the fluid-to-beam coupling assumptions will be demonstrated in Section \ref{sec:examples}. 
\end{remark}

\section{Numerical Examples}
\label{sec:examples}

The following numerical examples are chosen to illustrate the capabilities of the proposed fluid-beam interaction approach. To this end, this section focuses on the special cases of very stiff slender bodies and very light fibers in order to analyze basic properties of the one-way coupling cases. If not stated differently, all simulations include torsionfree beam elements. For all examples, the fluid is assumed to be initially at rest. All models are set up using the pre-processor MeshPy \cite{MeshPyWebsite}, and the simulations are performed with the in-house multi-physics research code BACI \cite{baci}.

\subsection{Rigid Obstacle Immersed in a Fluid Channel}
\label{subsec:obstacle}

    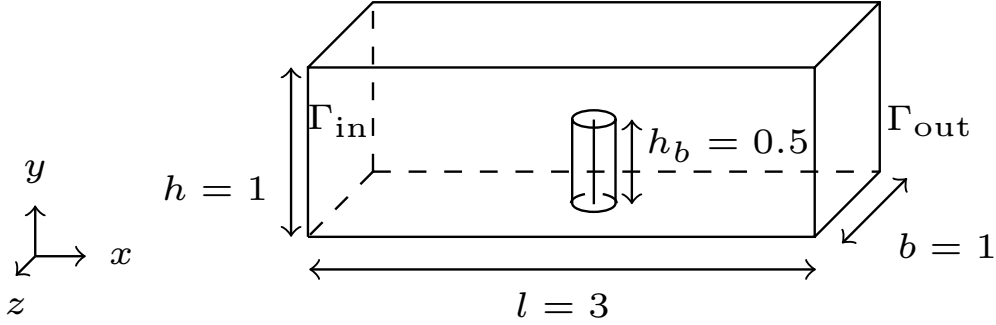
\begin{figure}[!h]
    	\begin{center}
    		\resizebox{0.9\textwidth}{!}{



\begin{tikzpicture}

\fill[white,opacity=.5] (0,0,0)-- (3,0,0) -- (3,1,0)  -- (0,1,0) --cycle;
\fill[white,opacity=.5] (0,0,1)-- (3,0,1) -- (3,1,1)  -- (0,1,1) --cycle;
\fill[white,opacity=.5] (0,1,0)-- (0,1,1) -- (3,1,1) -- (3,1,0)--cycle;
\fill[white,opacity=.5] (0,0,0)-- (0,0,1) -- (3,0,1) -- (3,0,0)--cycle; 
\draw[] (0,0,1) -- (3,0,1) -- (3,1,1) --(0,1,1) --(0,0,1)
(3,0,1) -- (3,0,0)  -- (3,1,0) --(0,1,0) -- (0,1,1)    
(3,1,1) -- (3,1,0);
\draw[dashed] (0,0,0) -- (0,0,1) (0,0,0)-- (3,0,0) (0,0,0)-- (0,1,0);

\draw[<->] (0.2,0,1.5)-- (3.2,0,1.5) node[pos=.5, below] {\tiny$l=3$};
\draw[<->] (3.2,0,0.1)-- (3.2,0,1.1) node[pos=.5, anchor= north west] {\tiny$b=1$};
\draw[<->] (-0.1,0,1)-- (-0.1,1,1) node[pos=.5, anchor=north east] {\tiny$h=1$};

\node (A) [cylinder,draw=black,aspect=0.4,
minimum height=0.6cm,minimum width=0.000001cm,
shape border rotate=90] at (1.5, 0.2, 0.5 ) {};

\draw[dashed]
let \p1 = ($ (A.after bottom) - (A.before bottom) $),
\n1 = {0.5*veclen(\x1,\y1)},
\p2 = ($ (A.bottom) - (A.after bottom)!.5!(A.before bottom) $),
\n2 = {veclen(\x2,\y2)}
in
(A.before bottom) arc [start angle=0, end angle=180,
x radius=\n1, y radius=\n2];

\draw (1.5,0, 0.5) -- (1.5, 0.5 ,0.5);

\draw[<->] (1.725,0, 0.5) -- (1.725, 0.5 ,0.5) node[pos=.7, anchor = west] {\hspace{-0.05cm}\tiny$h_b=0.5$};

\node at (-0.2,0.3) {\tiny{$\Gamma_\text{in}$}};
\node at (3.3,0.3) {\tiny{$\Gamma_\text{out}$}};

\draw[->] (-2.5, -1, -1.3) -- (-2.2, -1,-1.3) node[anchor=west] {\tiny$x$};
\draw[->] (-2.5, -1, -1.3) -- (-2.5, -0.7, -1.3) node[above] {\tiny$y$};
\draw[->] (-2.5, -1, -1.3) -- (-2.5, -1, -1) node[below] {\tiny$z$};

\end{tikzpicture}

}
    		\caption{Rigid obstacle immersed in a fluid channel}
    		\label{fig:obstaclesetup}
    	\end{center}
    \end{figure}
    
The purpose of the first example is twofold: its simplicity allows to easily compute the solution of a full 3D simulation to validate the proposed mixed-dimensional coupling approach, and to study the convergence behavior of the resulting numerical error with respect to uniform mesh refinement. Secondly, the example is used to illustrate the dependence of the constraint violation within the fluid partition on the chosen penalty parameter. While numerical constraint enforcement via the penalty method is common practice in constrained optimization problems as well as applications such as contact dynamics, a use within computational fluid dynamics as presented within this work is not as common. This novelty warrants a closer look at the effect of this choice of constraint enforcement method.

Figure \ref{fig:obstaclesetup} shows the configuration of a rigid obstacle immersed in a mono-directional fluid flow. The time-dependent inflow velocity in x-direction $v_{in}=0.5\cdot\left(1-\cos\left(10\pi t\right)\right)$ for $t\in \left[0, 0.1\right]$, and $v_{in}=1$ for $t>0.1$, is prescribed on the surface $\Gamma_{in}$. On the surface $\Gamma_{out}$, a zero-traction condition is applied and perfect-sliding conditions are enforced on all other surfaces. The rigid beam obstacle is placed in the middle of the lower channel surface and is assumed to not be affected by the surrounding fluid. This allows to consider only one-way coupling from the rigid beam onto the fluid. The fluid is assumed to have a density of $\rho = 1$ and a dynamic viscosity of $\mu = 0.004$. The fluid domain is discretized by $96 \times 32 \times 32$ stabilized 8-noded hexahedral finite elements with equal-order interpolation, and the evolution in time is solved by the Crank-Nicolson method, i.e. the one-step-$\theta$ method with $\theta=\dfrac{1}{2}$, with time step size $\Delta t = 10^{-3}.$ Note that for the proposed approach therefore neither the radius of the beam obstacle nor its material properties enter the simulation. Instead, only the effect of the applied Dirichlet conditions on the fluid solution in dependence on the penalty parameter shall be analyzed.

\begin{figure}[!h]
	\begin{subfigure}[b]{0.625\textwidth}
		\includegraphics[width=\textwidth]{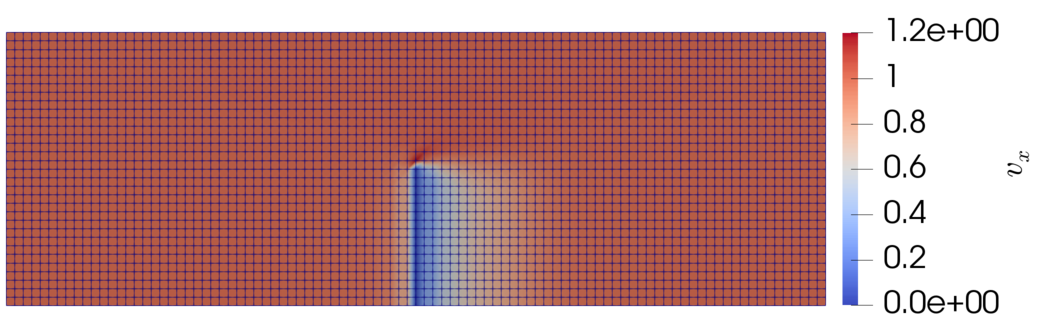}
		\subcaption{Fluid velocity $v_x$ in channel direction in the plane of the rigid obstacle}
		\label{fig:obstacle_vel}
	\end{subfigure}
	\hspace{1cm}
	\begin{subfigure}[b]{0.3\textwidth}
		\includegraphics[width=\textwidth]{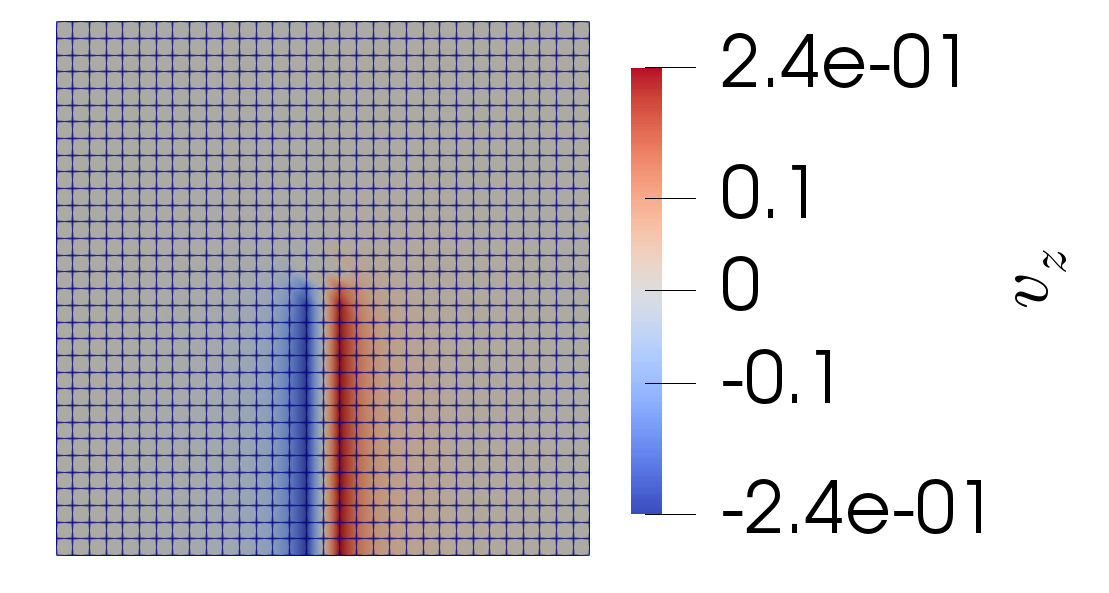}
        \subcaption{Fluid velocity $v_z$ around the obstacle}
		\label{fig:obstacle_zvel}
	\end{subfigure}
	\caption{Visualization of the effect of the obstacle on the fluid velocity in the channel}
	\label{fig:obstacle_visualization}
\end{figure}

\begin{figure}[!h]
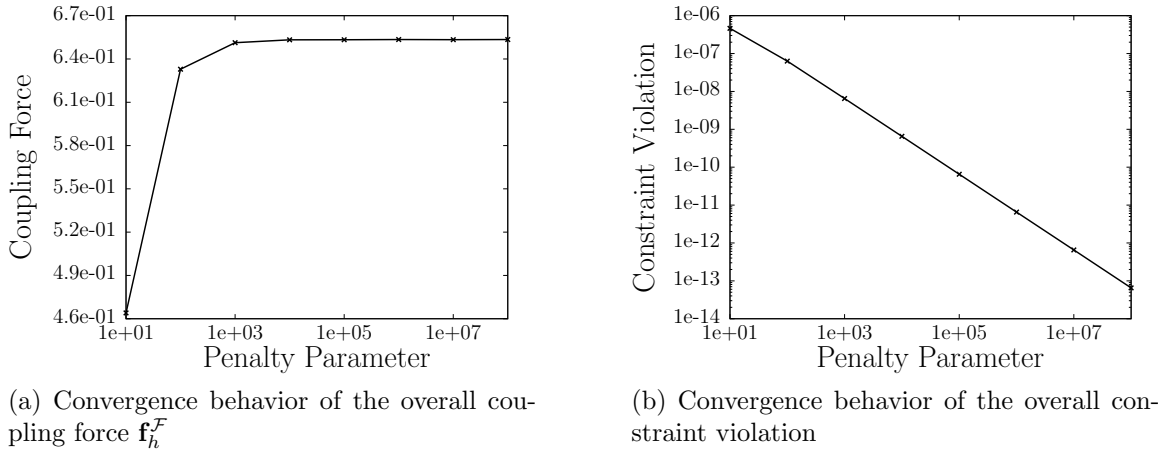

	\begin{subfigure}[b]{0.45\textwidth}
		\resizebox{\textwidth}{!}{\input{pictures/hindernis_lowre/force.tex}}
		\subcaption{Convergence behavior of the overall coupling force $\mathbf{f}_h^\mathcal{F}$}
		\label{Fig:constraint_force}
	\end{subfigure}
	\hspace{1cm}
	\begin{subfigure}[b]{0.45\textwidth}
		\resizebox{\textwidth}{!}{\input{pictures/hindernis_lowre/violation.tex}}
		\subcaption{Convergence behavior of the overall constraint violation}
		\label{Fig:constraint_violation}
	\end{subfigure}
	\caption{Convergence behavior of the overall constraint violation with respect to the penalty parameter}
	\label{Fig:penalty_convergence}
\end{figure}

The obstacle is expected to decelerate the fluid in its vicinity to a resting state, and thereby redirect the fluid flow to the sides and to the top. For a penalty parameter $\epsilon=10^4$, simulation results illustrating this behavior are shown in Figure \ref{fig:obstacle_visualization}. The deceleration in x-direction as well as the deflection of the flow velocity in z-direction is clearly visible. In the regarded case of a rigid beam at rest, the constraint violation after time step $n+1$ is measured by calculating the $L_2$-norm of $\mathbf{K}_\mathcal{FF}^{n+1}\mathbf{v}_h^{f, n+1}$. Figure \ref{Fig:constraint_violation} shows the dependence of the constraint violation on the penalty parameter at time $t=0.5$ when a steady state solution is reached.

In addition, Figure \ref{Fig:constraint_force} allows a closer look at the $L_2$-norm of the computed penalty force $\mathbf{f}_h^{\mathcal{F}, n+1}=\epsilon\mathbf{K}^{n+1}_\mathcal{FF}\mathbf{v}_h^{f, n+1}$ acting on the fluid at time $t=0.5$.
In Section \ref{sec:algo}, it was argued, that this penalty force will take on the amount of force necessary to fulfill the given FBI constraint, and thus can be interpreted as the interface force acting onto the fluid.
For this hypothesis to be valid, the penalty force needs to be practically independent of the chosen value of the penalty parameter, or equivalently, linear convergence of the constraint violation $\mathbf{K}^{n+1}_\mathcal{FF}\mathbf{v}_h^{f, n+1}$ towards 0 is expected as the penalty parameter increases.
Figure \ref{Fig:penalty_convergence} suggests that both assumptions hold true for sufficiently large values of the penalty parameter. This is also an important basis for further work towards a fully coupled FBI framework, where the position of the beam centerline, and thus the geometry of the coupled problem, as well as the beam velocity, depend on the penalty force.

\begin{figure}[!h]
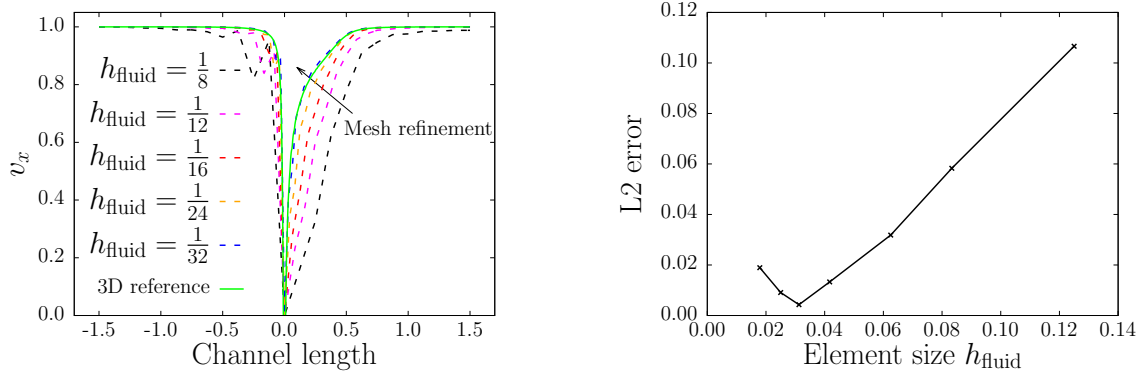

    	\begin{subfigure}[b]{0.45\textwidth}
		\resizebox{\textwidth}{!}{\input{pictures/hindernis_lowre/velocity_profile_small.tex}}
        \subcaption{Fluid velocity profile $v_x$ over the channel length for different fluid element sizes $h_\text{fluid}$ as well as a finely resolved 3D reference solution}
\label{fig:obstacle_vel_profile}
    \end{subfigure}
	\hspace{1cm}
	\begin{subfigure}[b]{0.45\textwidth}
		\resizebox{\textwidth}{!}{\input{pictures/hindernis_lowre/spacial_convergence.tex}}
		\subcaption{L2 error of the fluid velocity in x-direction over the fluid element size $h_\text{fluid}$ computed using a 3D CFD solution as reference}
		\label{fig:obstacle_mesh_convergence}
	\end{subfigure}
	\caption{Dependence of the fluid velocity on the mesh resolution}
	\label{fig:obstacle_meshconv}
\end{figure}

Next, the behavior of the 1D-3D coupling solution with respect to the fluid mesh resolution shall be analyzed. Figure \ref{fig:obstacle_vel_profile} shows the space-dependent steady state solution of $\mathbf{v}_x$ along the x-axis for different mesh resolutions. Here, the reference solution is computed using a full 3D computational fluid dynamics (CFD) simulation, in which the 3-dimensional beam domain is cut out of the fluid domain, and no-slip boundary conditions are enforced on the newly generated beam surface. The fluid field is discretized using 641.928 DoFs.
This 3D reference solution is used to compute the L2 error of the shown 1-dimensional velocity profile plotted over the fluid element size $h_\text{fluid}$ in Figure \ref{fig:obstacle_mesh_convergence}.

Examining the fluid velocity profile over the channel length plotted in Figure \ref{fig:obstacle_vel_profile}, the fluid flow is slowed down by the beam, while the flow is undisturbed far from the obstacle as expected. The kink in the FBI solutions just before the beam stems from the neglect of capturing the exact behavior of the pressure solution at the beam. As investigated by Baaijens in \cite{Baaijens2001}, the pressure would exhibit a jump at the beam, which cannot be represented within the discretization space of continuous piece-wise linear functions. Using appropriate discretization spaces allowing for pressure jumps on element boundaries as in \cite{Baaijens2001} or even enhancing the space as in the extended finite element method in \cite{schott2015} can potentially solve this problem. We deliberately refrain from such advanced fluid discretizations, since we are targeting an efficient solver for macroscopic effects in FBI, not a fine-scale resolution in the vicinity of the coupling interface. For now, note that refinement of the mesh near the obstacle can smooth these kinds of kinks.

Further, it becomes evident that the fluid profile computed with the FBI method at a fluid element size of $h_\text{fluid}=\dfrac{1}{32}$ matches the 3D reference solution very well. As a more quantitative analysis, the L2 error of the 1D profiles with respect to the 3D CFD solution is shown in Figure \ref{fig:obstacle_mesh_convergence}. The 1D-3D coupling approach exhibits a linear convergence behavior with respect to uniform mesh refinement, which is expected due to the fact that the beam radius does not enter the simulation, and instead, the area in which the fluid is slowed down is scaled by the fluid element size. This convergence behavior as well as the error values are in line with the results for general academic mixed-dimensional models reported in \cite{kerfriden2020, zunino2020, laurino2020}, none of which exceeded linear convergence in the primary variable of the 3-dimensional field. It can also be seen, that for the proposed FBI approach, further mesh refinement of the fluid background mesh will lead to a narrowing of the affected fluid area and, thus, a growing error with respect to the 3D reference solution as evident in Figure \ref{fig:obstacle_mesh_convergence}.
This convergence behavior closely resembles the one that has been observed and discussed for a similar approach applied to mixed-dimensional solid-beam coupling in \cite{steinbrecher2020}.
Such a behavior is to be expected, as the proposed 1D-3D coupling approach is only valid under the assumptions of relatively small beam radii compared to the fluid element size as elaborated in Remark \ref{remark:element_size}. For problem setups, in which the model assumptions are violated, no further convergence can be expected and a different modeling technique should be chosen.

In conclusion, this example demonstrates that the FBI solution converges towards the reference solution within a spectrum of fluid element sizes, for which the model assumptions, as discussed in Remark \ref{remark:element_size}, are fulfilled. While recovery of optimal convergence for 1D-3D coupling approaches is still a worthwhile topic of ongoing research, the matching of the solutions obtained with mixed-dimensional and fully resolved models satisfies the goals of this work and validates the general applicability of the proposed FBI model.

\subsection{Rotating Lattice}
\label{subsec:mixer}

\begin{figure}[!h]
	\begin{subfigure}[b]{0.45\textwidth}
		\begin{center}
		\resizebox{\textwidth}{!}{\hspace{0.5cm} \begin{tikzpicture}

\draw[thick] (-0.83333, -0.83333) -- (-0.83333, 0.83333);
\draw[thick] (-0.5, -0.83333) -- (-0.5, 0.83333);
\draw[thick] (-0.16666, -0.83333) -- (-0.16666, 0.83333);
\draw[thick] (0.16666, -0.83333) -- (0.16666, 0.83333);
\draw[thick] (0.5, -0.83333) -- (0.5, 0.83333);
\draw[thick] (0.83333, -0.83333) -- (0.83333, 0.83333);

\draw[thick] (-0.83333, -0.83333) -- (0.83333, -0.83333);
\draw[thick] (-0.83333, -0.5) -- (0.83333, -0.5);
\draw[thick] (-0.83333, -0.16666) -- (0.83333, -0.16666);
\draw[thick] (-0.83333, 0.16666) -- (0.83333, 0.16666);
\draw[thick] (-0.83333, 0.5) -- (0.83333, 0.5);
\draw[thick] (-0.83333, 0.83333) -- (0.83333, 0.83333);

\draw[<->] (1, -0.83333) -- (1, 0.83333) node[pos=0.5, anchor=west] {\tiny{h=$1.\bar{6}$}};
\draw[<->] (-0.83333, 1) -- (0.83333, 1) node[pos=0.5, anchor=south] {\tiny{b=$1.\bar{6}$}};

\draw[->] (-1.5, -1) -- (-1.2, -1) node[anchor=west]{\tiny{$y$}};
\draw[->] (-1.5, -1) -- (-1.5, -0.7) node[anchor=south]{\tiny{$z$}};

\end{tikzpicture}}
		\subcaption{Geometry of the beam lattice}
		\label{Fig:lattice}
		\end{center}
	\end{subfigure}
	\hspace{1cm}
	\begin{subfigure}[b]{0.45\textwidth}
		\begin{center}
			\resizebox{0.8\textwidth}{!}{\begin{tikzpicture}

\draw[thick] (0, 0) circle (1);
\draw[thick] (0, -0.83333) -- (0, 0.83333);
\draw[<->] (-0.1, -0.83333) -- (-0.1, 0.83333) node[pos=0.5, anchor=east] {\tiny{b=$1.\bar{6}$}};
\draw[<->] (-1.2, -1) -- (-1.2, 1) node[pos=0.5, anchor=east] {\tiny{d=$2$}};

\draw [densely dashed, ->] (0.1, 0.7) to [out=-10, in=10] ( 0.1, -0.7);
\node at (0.75, 0) {\tiny{$\mathbf{\Delta \bar{r}}$}};

\draw[->] (-2, -1.1) -- (-1.7, -1.1) node[anchor=west]{\tiny{$x$}};
\draw[->] (-2, -1.1) -- (-2, -0.8) node[anchor=south]{\tiny{$y$}};

\end{tikzpicture}}
			\subcaption{Top view of the immersed lattice}
			\label{Fig:mixer_sketch}
		\end{center}
	\end{subfigure}
	\caption{Configuration of the rotating lattice immersed in a cylindrical fluid tank}
\end{figure}
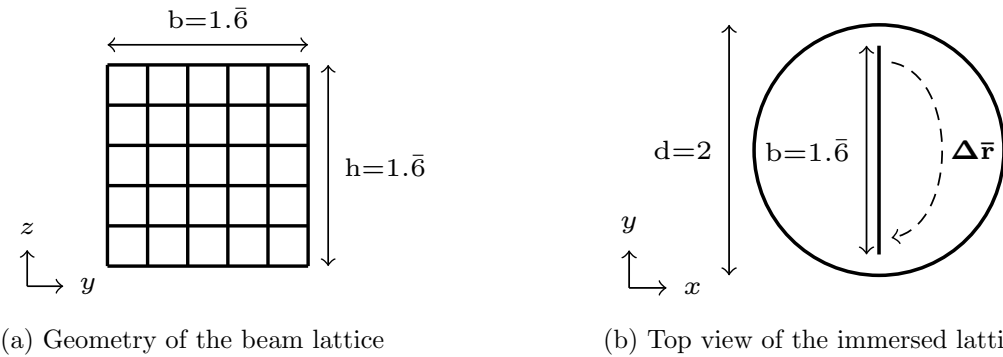

\begin{figure}[h!]
    \begin{subfigure}[b]{0.45\textwidth}
        \resizebox{\textwidth}{!}{\input{pictures/mixer/mixer_profile.tex}}
        \subcaption{Fluid velocity Magnitude over the beam lattice in radial direction}
        \label{Fig:mixer_profile}
    \end{subfigure}
    \hspace{1cm}
    \begin{subfigure}[b]{0.45\textwidth}
        \includegraphics[width=\textwidth]{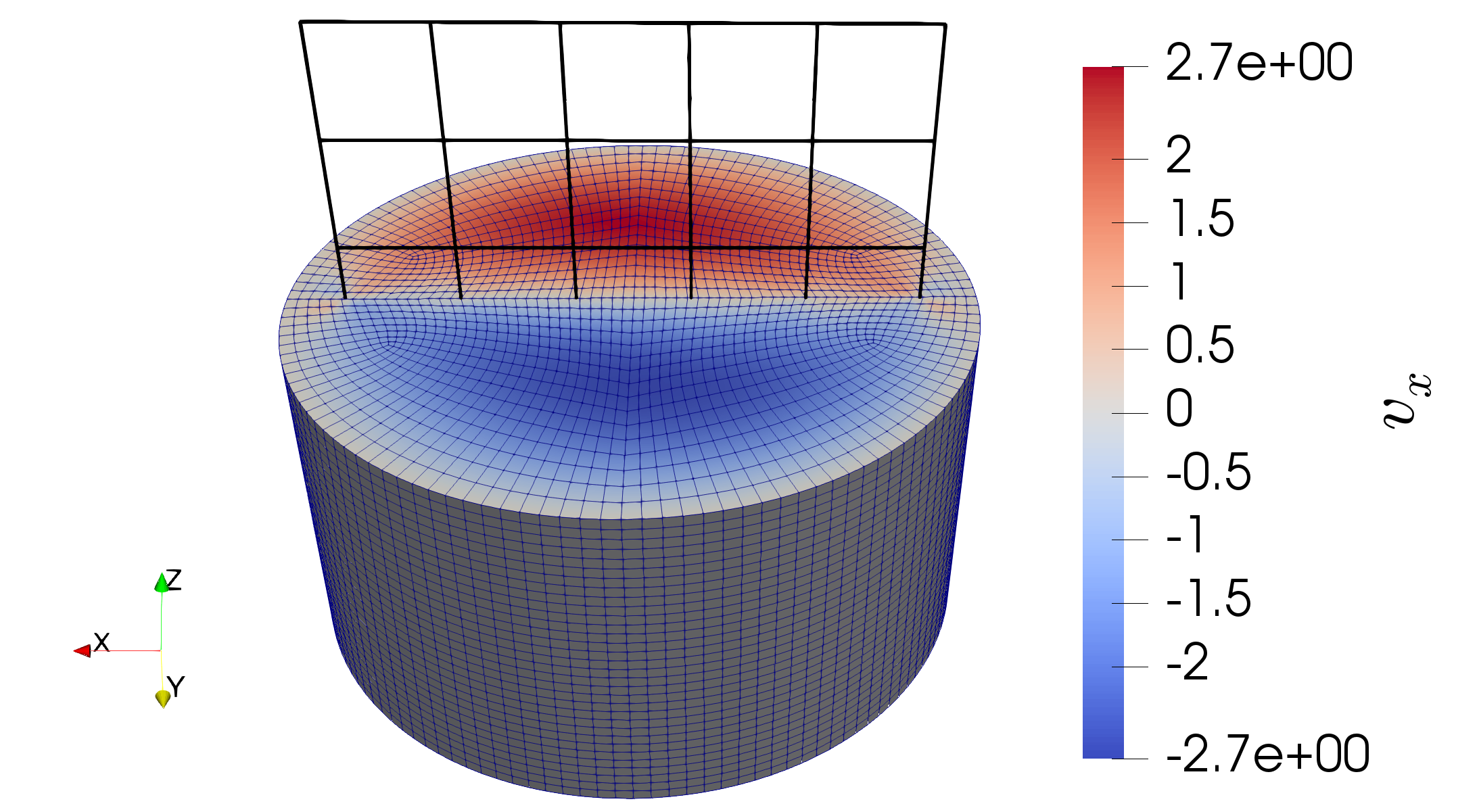}
        \subcaption{Fluid velocity in x-direction after three quarters of a rotation of the beam lattice}
        \label{Fig:mixer_picture}
    \end{subfigure}
    \caption{Fluid velocity profile induced by the moving beam lattice}
    \label{Fig:mixer_solution}
\end{figure}

After analyzing the behavior of the constraint violation itself, we now want to observe the global effects of slender bodies on fluid flow as well as the proposed method's robustness under large displacements. To this end, we will regard a rigid lattice of dimension $1.\bar{6}\times 1.\bar{6}$ made up of beams, as depicted in Figure \ref{Fig:lattice}, immersed in a cylindric tank filled with fluid at rest. The cylinder has a radius of 1, and a height of 2 in order to encompass the entire beam lattice. Furthermore, all surfaces are modeled with no-slip boundary conditions. The fluid is assumed to have a density of 1.0 and a dynamic viscosity of 4, while the beams are again assumed to not be affected by the fluid itself. The lattice starts rotating around the vertical axis of the cylinder as depicted in Figure \ref{Fig:mixer_sketch}. The movement is described by the rigid body motion

\begin{equation}
\mathbf{\Delta \bar{r}}=
\left(
\begin{array}{c}
\cos\left(2\cdot\pi\cdot \bar{v}\right)\cdot x - \sin\left(2\cdot \pi \cdot \bar{v}\right) \\
\sin\left(2\cdot\pi\cdot \bar{v}\right)\cdot x + \cos\left(2\cdot \pi \cdot \bar{v}\right)
\end{array}\right),
\end{equation}
with the time-dependent scaling factor $\bar{v}=0.5\cdot \left(1-\cos\left(2\pi t\right)\right)$ for $t\in\left[0,0.5\right]$, and $\bar{v}=1$ for $t \geq 0.5$. For this example, the time step is chosen as $\Delta t = 10^{-3}$, the evolution in time is discretized with the Backward Euler time stepping scheme, and the FBI constraint is enforced with a penalty parameter of $\epsilon=100$. The entire fluid domain is discretized with $114,688$ hexahedral finite elements arranged into $64$ layers along the height of the cylinder. The fluid mesh is depicted in Figure \ref{Fig:mixer_picture}.

It is expected that the beam lattice incites the fluid within the cylinder to start rotating. Herein, the fluid velocity in the vicinity of the beam matches the rigid body motion up to a penalty constraint violation, while the velocity further away from the rotator is indirectly accelerated by the surrounding fluid.

\begin{figure}[h!]
	\begin{subfigure}[b]{0.45\textwidth}
		\includegraphics[width=\textwidth]{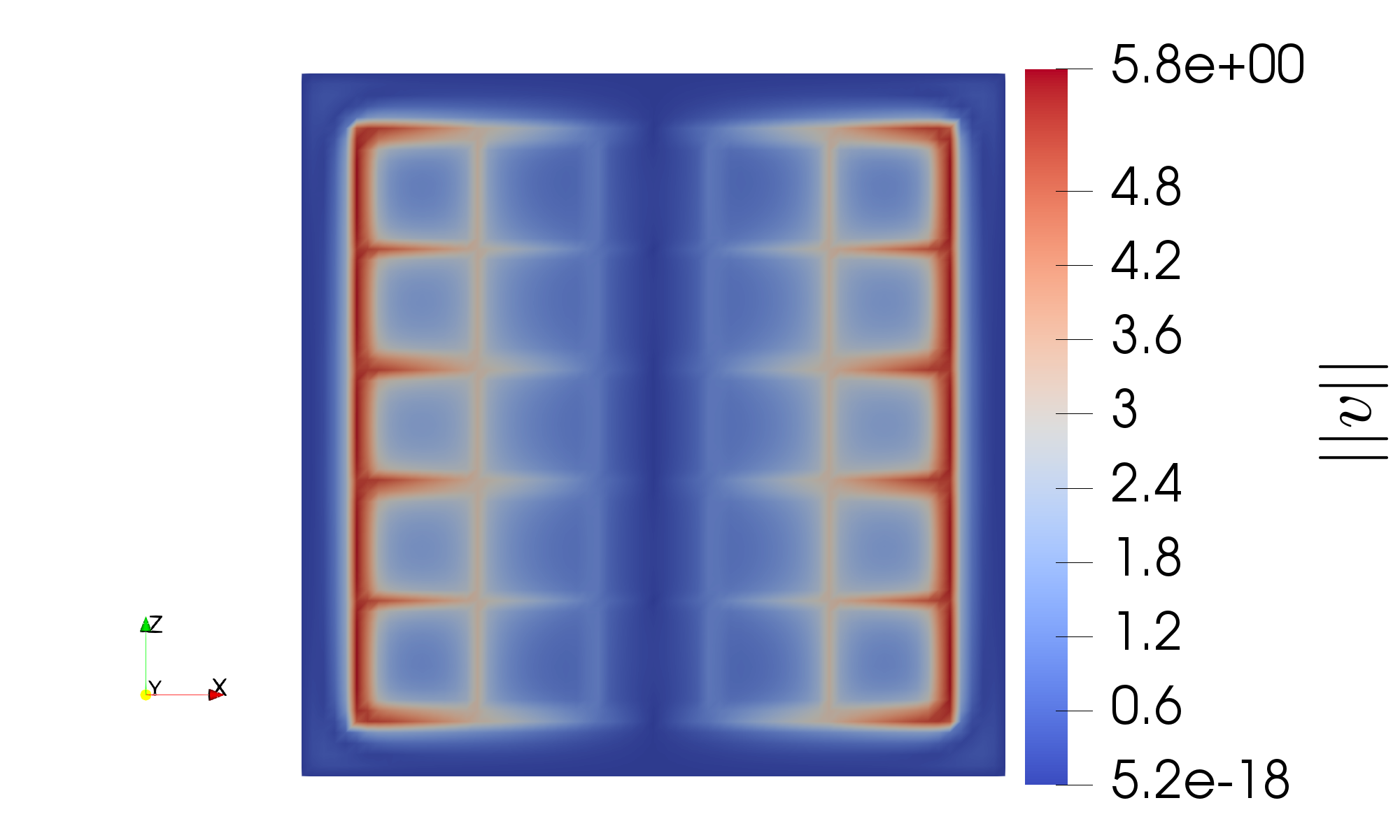}
		\subcaption{Magnitude of the fluid velocity in the vicinity of the beam lattices induced by the prescribed lattice movement}
		\label{Fig:mixer_picture_height}
	\end{subfigure}
		\hspace{1cm}
	\begin{subfigure}[b]{0.45\textwidth}
		\resizebox{\textwidth}{!}{\input{pictures/mixer/mixer_height_profile.tex}}
		\subcaption{Fluid velocity in y-direction over the lattice height in comparison to the prescribed rigid body motion}
		\label{Fig:mixer_profile_height}
	\end{subfigure}
	\caption{Fluid velocity profile induced by the moving beam lattice along the height of the fluid cylinder}
	\label{Fig:mixer_height_solution}
\end{figure}
				
Figures \ref{Fig:mixer_height_solution} and \ref{Fig:mixer_solution} depict the fluid solution after three quarters of a rotation of the lattice at time $t=0.837$. Note that in the regarded time step, the velocity of the lattice in x-direction is negligible.

Figure \ref{Fig:mixer_picture_height} shows that the absolute velocity $\left\|\mathbf{v}\right\|_2$ follows the lattice structure of the beam geometry. Figure \ref{Fig:mixer_profile_height} offers a closer look at the velocity measured over the cylinder height at the coordinate $x=0.7$ and $y=0$. It can be seen that the local velocity extrema of the fluid velocity $v_y$ (black line) do not exactly match the beam positions (vertical lines). Instead, the fluid velocity in the vicinity of the beam struts is even a little higher than the prescribed rigid body velocity (blue line). This can be explained by the fact that, within the proposed coupling approach, the fluid solution is not enriched by additional shape functions to model gradient jumps within elements as is the case for extended finite elements as used in \cite{schott2015}. Thus, no sudden change in the fluid velocity within an element can be represented. The local extrema of the flow velocity thus fall to the finite element node closest to the beam, as seen in \ref{Fig:mixer_profile_height}, and lead to a discretization error near the interface, caused by the reduced complexity of the proposed approach.

Figure \ref{Fig:mixer_profile} shows the fluid velocity profile $v_y$, and thus the total velocity, in radial direction along the x-axis cutting the beam lattice in a top and a bottom half. As expected from a physical point of view, it can be seen that the fluid velocity matches the rotation velocity very closely in the vicinity of the beam lattice while it lags behind within the holes of the mesh. Clearly, for this case of a less steep evolution of the fluid solution than in Figure \ref{Fig:mixer_profile_height}, the discretization error due to missing enrichments of the underlying discretization space is less pronounced.

Figure \ref{Fig:mixer_picture} shows the fluid mesh, which was traversed via different cut scenarios of the overall beam lattice with the background mesh within each time step, until the presented state was reached. In the present case, the fluid mesh contains $114,688$ fluid elements, while the lattice is discretized using $1,200$ beam elements in order to force the the creation of numerous different integration segments by the segmentation procedure visualized in Figure \ref{fig:segmentation}. In this case, application of the segmentation procedure leads to the creation of a number of different integration segments between $1,597$ and $1,940$ for each time step. Here, the exact number of segments for each time step depends on the position of the lattice relative to the background mesh and, thus, varies rather considerably over the course of the simulation due to the large lattice displacement and the numerous different cut scenarios stemming from the unsymmetric meshing of the fluid tank. Therefore, the regarded example also serves as a preliminary validation of the robustness of the applied segmentation procedure shown in Figure \ref{fig:segmentation} with respect to varying intersection scenarios of the embedded with the background mesh.

\subsection{Light Fibers in Fluid Flow}
\label{subsec:fibres}

Having analyzed the one-way coupled case of rigid beams affecting fluid flow, this section is meant to investigate the effect of the penalty parameter on the coupling of a light fiber being transported by a fluid.

In this example, the fluid is again assumed to be contained in a hexahedron of dimensions $1\times1\times3$ as introduced in Section \ref{subsec:obstacle}. Only now all DoFs in channel direction are set to be free and a velocity of zero is prescribed in both other directions, leading to a pseudo one-dimensional setup. In order to analyze the time-dependent behavior of the beam in dependence on the penalty parameter, an oscillatory inflow velocity

\begin{equation}
	v_{in} = 0.5\cdot\left(1-\cos\left(\pi\cdot 10\cdot t\right)\right)
\end{equation}
in channel direction, is prescribed at the channel inlet.

The behavior of an immersed fiber with a density $\rho=1$, a length $l=0.5$, a cross-sectional area $A=0.166$, and accordingly a mass of $m=0.0880$ is regarded. Note that the other material properties of the beam do not play a role in this example, since the fluid flow is set up to be constant along the beam length such that the results can be analyzed as an immersed mass in dependence on the penalty parameter. The time step size is chosen as $\Delta t=0.01$.

\begin{figure}[h!]
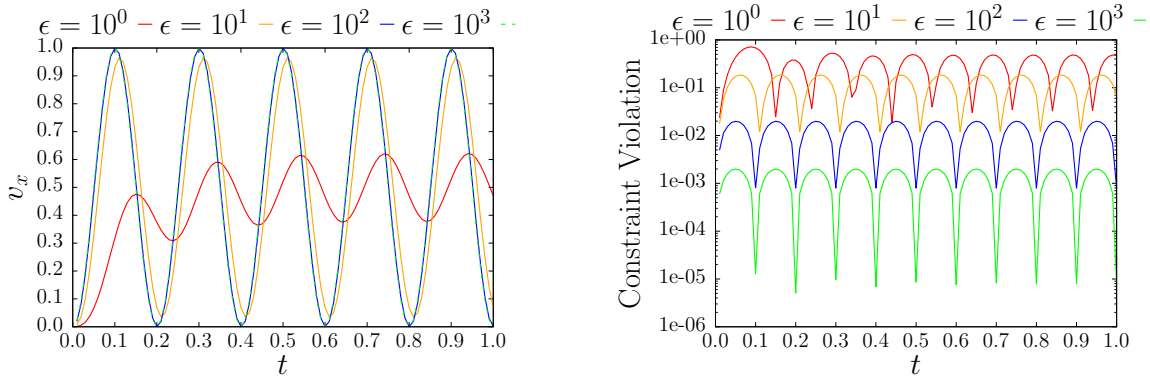

    \begin{center}
        \begin{subfigure}{0.45\textwidth}
            \resizebox{\textwidth}{!}{\input{pictures/flow_oscillating/beam_velocity_oscilating.tex}}
            \caption{Velocity of a rigid slender body immersed in a fluid channel for an oscillatory fluid inflow profile}
            \label{Fig:beam_velocity}
        \end{subfigure}
        \hspace{1cm}
        \begin{subfigure}{0.45\textwidth}
            \resizebox{\textwidth}{!}{\input{pictures/flow_oscillating/constraintviolation_over_penalty.tex}}
            \caption{Absolute difference between the beam velocity and the prescribed fluid velocity over time}
            \label{Fig:beam_error}
        \end{subfigure}
    \end{center}
    \caption{Behavior of an immersed freely moving beam in dependence of the penalty parameter}
    \label{Fig:beam_penalty}
\end{figure}

In the limit case of freely movable light fibers, the fiber is expected to be transported by the fluid exactly at the velocity of the fluid. Figure \ref{Fig:beam_velocity} shows velocity results for the beam in dependence on the penalty parameter. As argued in Remark \ref{remark:DN}, within the fluid-to-beam one-way coupling variant, the penalty force is used to introduce a weak Dirichlet constraint into the structure problem. As in the case of the beam-to-fluid coupling one-way coupling variant, we thus expect linear convergence of the coupling violation towards zero for sufficiently large penalty parameters. Figure \ref{Fig:beam_error} exhibits this expected behavior. In particular, this validates the treatment of the geometry-dependence by a fixed-point iteration as proposed in Figure \ref{sec:algo} instead of a full Newton method.
    
\subsection{Contacting Immersed Beams}
\label{subsec:contact}

\begin{figure}[h!]
	\begin{subfigure}[b]{0.48\textwidth}
		\includegraphics[width=\textwidth]{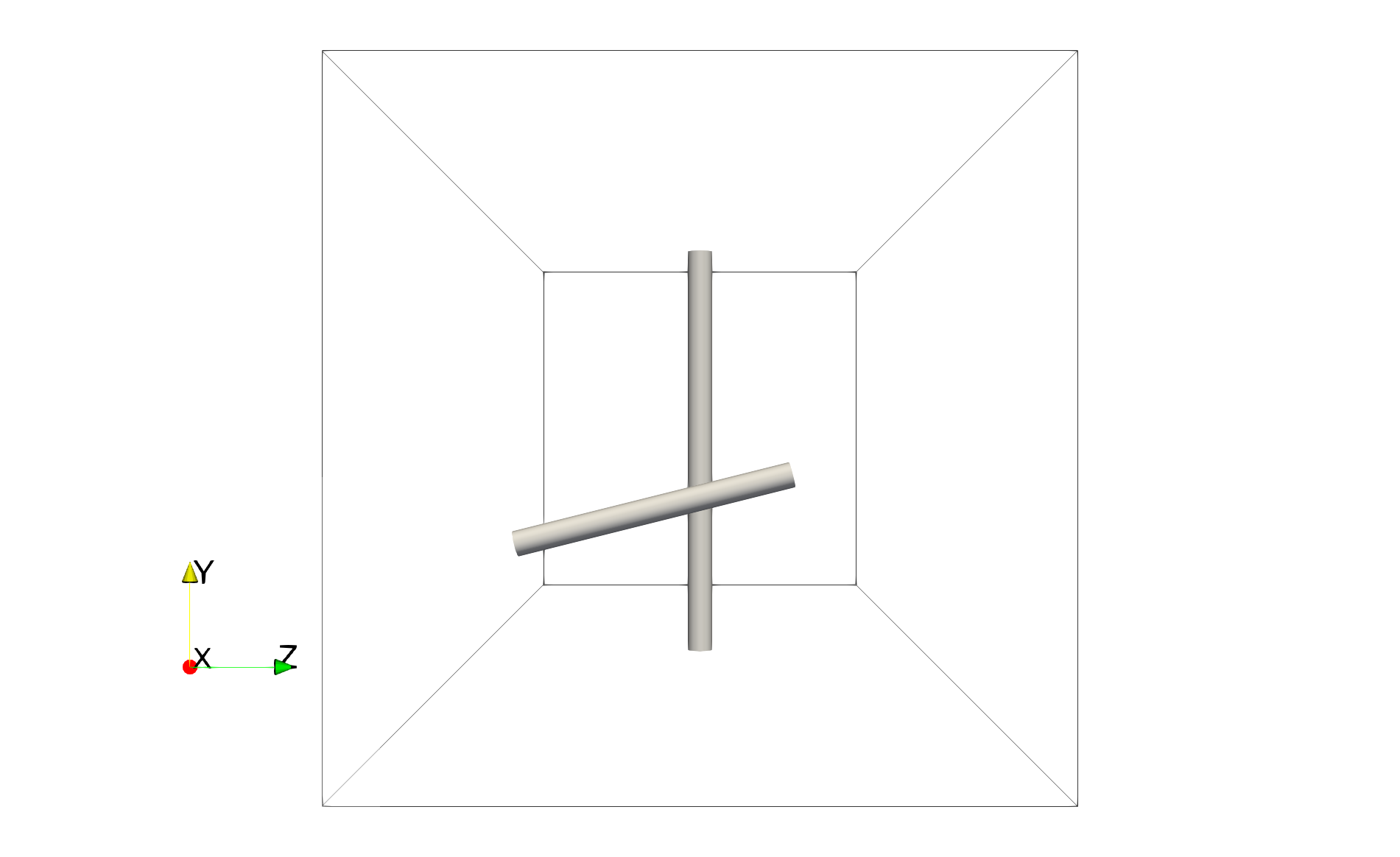}
		\subcaption{Setup of the geometry of two contacting beams}
		\label{Fig:contact_front}
	\end{subfigure}
	\hspace{1cm}
	\begin{subfigure}[b]{0.42\textwidth}
		\includegraphics[width=\textwidth]{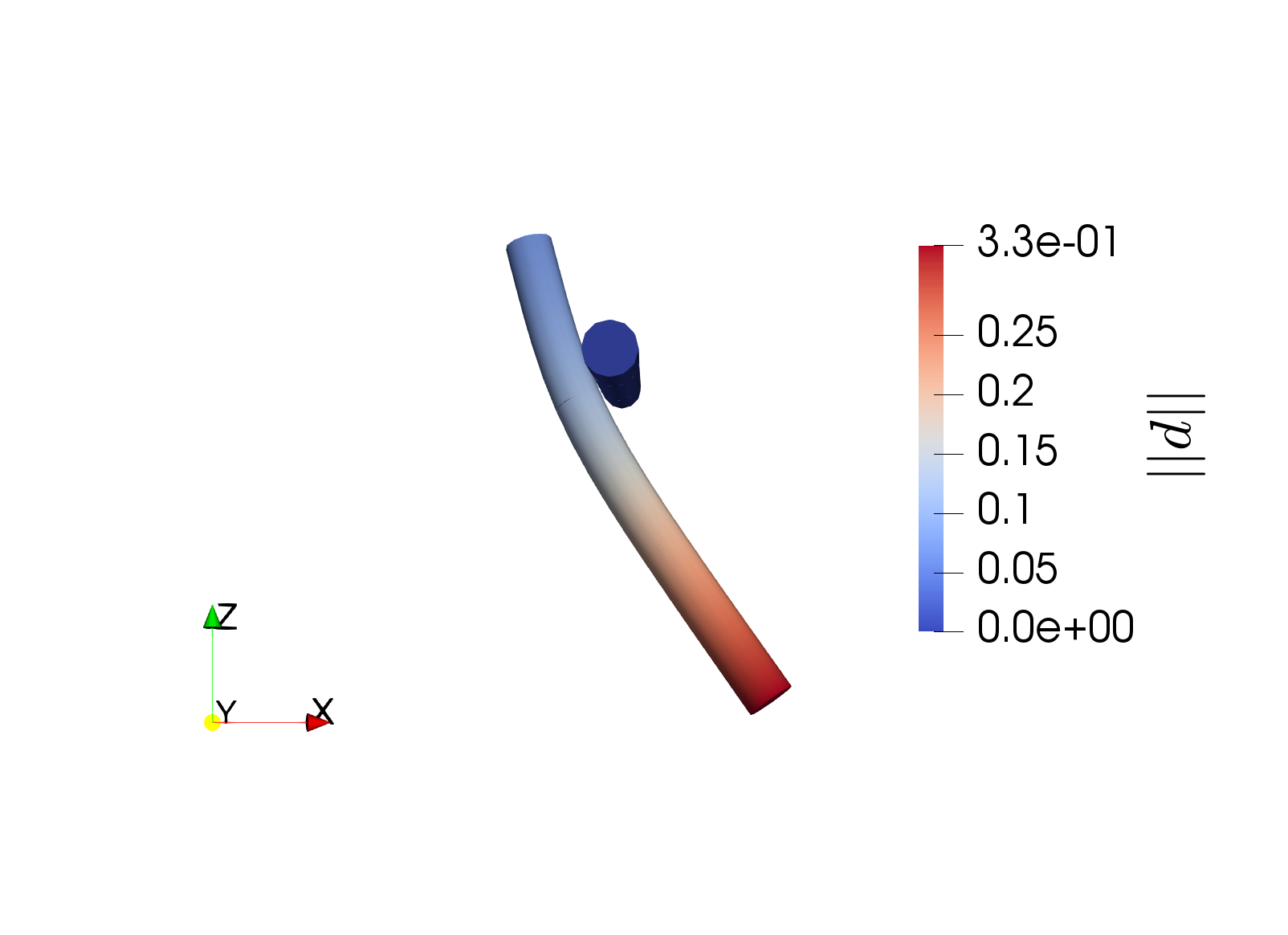}
		\subcaption{Free beam sliding along the fixed beam obstacle}
		\label{Fig:contact_top}
	\end{subfigure}
	\caption{Contacting fibers}
	\label{Fig:contact}
\end{figure}

Next, we want to study the applicability of the proposed approach as a simplified method to capture the effect of fluid-driven contact of multiple beams. In the case of fluid-driven contact, generally, multiple constraints act on the beam, namely the non-penetration contact constraint in addition to the kinematic FBI constraint. In the case of the fluid-to-beam one-way coupling variant, where the fluid is assumed to not be affected by the beam, in general it is not possible for both constraints to be perfectly fulfilled at the same time. Nevertheless, we will show the suitability of the proposed approach to model fluid-driven beam-to-beam contact under rather small violations of the one-way-coupling assumptions and moderate FBI penalty parameters, using a penalty contact method as introduced in \cite{Meier2016_contact}.

For this example, the fluid is again enclosed in a $1\times1\times3$ hexahedron, and the fluid at the inflow boundary is slowly accelerated to $v_x=3$ along the longitudinal channel direction. The first beam is secured at the middle of the bottom plane as in Figure \ref{fig:obstaclesetup}, but with a height of $0.9$, a Young's modulus $E=10^6$, a density $\rho=10^{3}$, and a cross-section area $A=0.25$. The second beam is freely flowing, has a Young's modulus of $E=10^5$, a density $\rho=10^{-3}$, a cross-sectional area $A=0.25$, and is positioned at an angle of approximately $105^\circ$, $0.15$ upstream from the first beam, as depicted in Figure \ref{Fig:contact_front}. Contact itself is modeled using tailored beam-to-beam contact formulation, as introduced in \cite{Meier2016_contact}, with the contact penalty parameter $10^6$, while the FBI penalty parameter is chosen as $1$.

\begin{figure}[t!]
    \begin{center}
        \resizebox{0.65\textwidth}{!}{\input{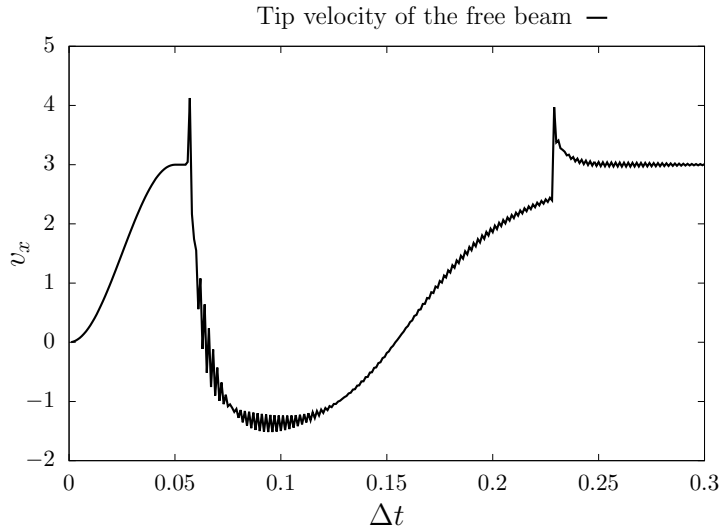}}
        \caption{Tip velocity in x-direction of a free contacting beam}
        \label{Fig:contact_velocity}
    \end{center}
\end{figure}

The free flowing beam is expected to be transported with the fluid velocity until the time of first contact. After coming into contact, the free beam will begin sliding along one side of the fixed beam obstacle, as depicted in Figure \ref{Fig:contact_top}, due to the geometrical unsymmetry of the problem setup.

Figure \ref{Fig:contact_velocity} shows the velocity of the tip of the free flowing beam, which is depicted as the right tip in Figure \ref{Fig:contact_front}. Besides the above described behavior, an incitation of the eigenfrequencies can be observed right after contact first occurs and after it ends again. However, these oscillations are promptly damped by the surrounding fluid. This is to show that the proposed fluid-to-beam one-way coupling algorithm can adequately capture the behavior of contacting beams and is thus also extensible to practical applications moderately violating the one-way coupling assumptions.

\subsection{Immersed Stent Geometry}
\label{stent}

\begin{figure}[h!]
	\begin{subfigure}[b]{0.45\textwidth}
		\includegraphics[width=\textwidth]{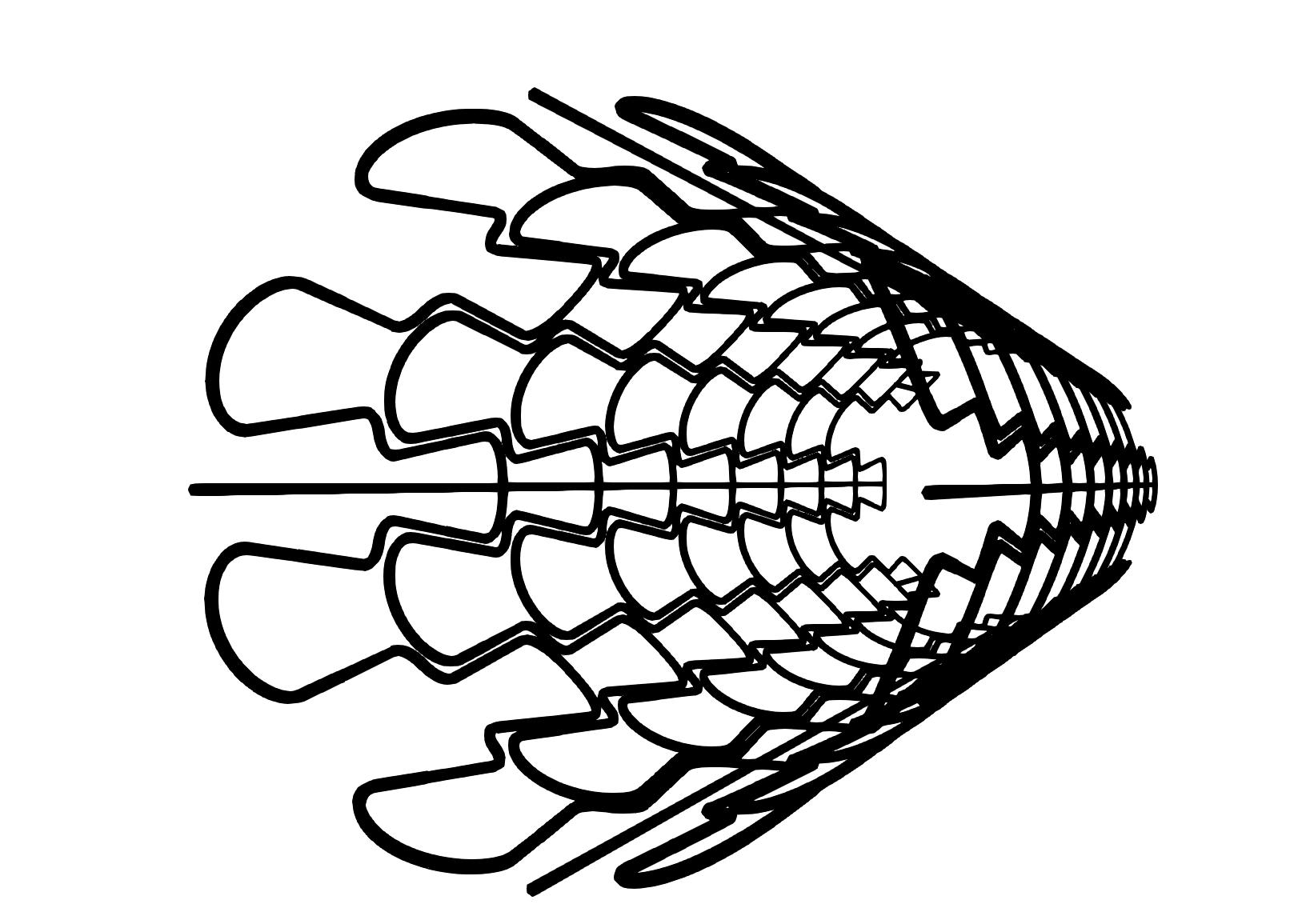}
		\subcaption{Beam centerline reference configuration}
		\label{Fig:stent_geometry}
	\end{subfigure}
	\hspace{1cm}
	\begin{subfigure}[b]{0.45\textwidth}
		\includegraphics[width=\textwidth]{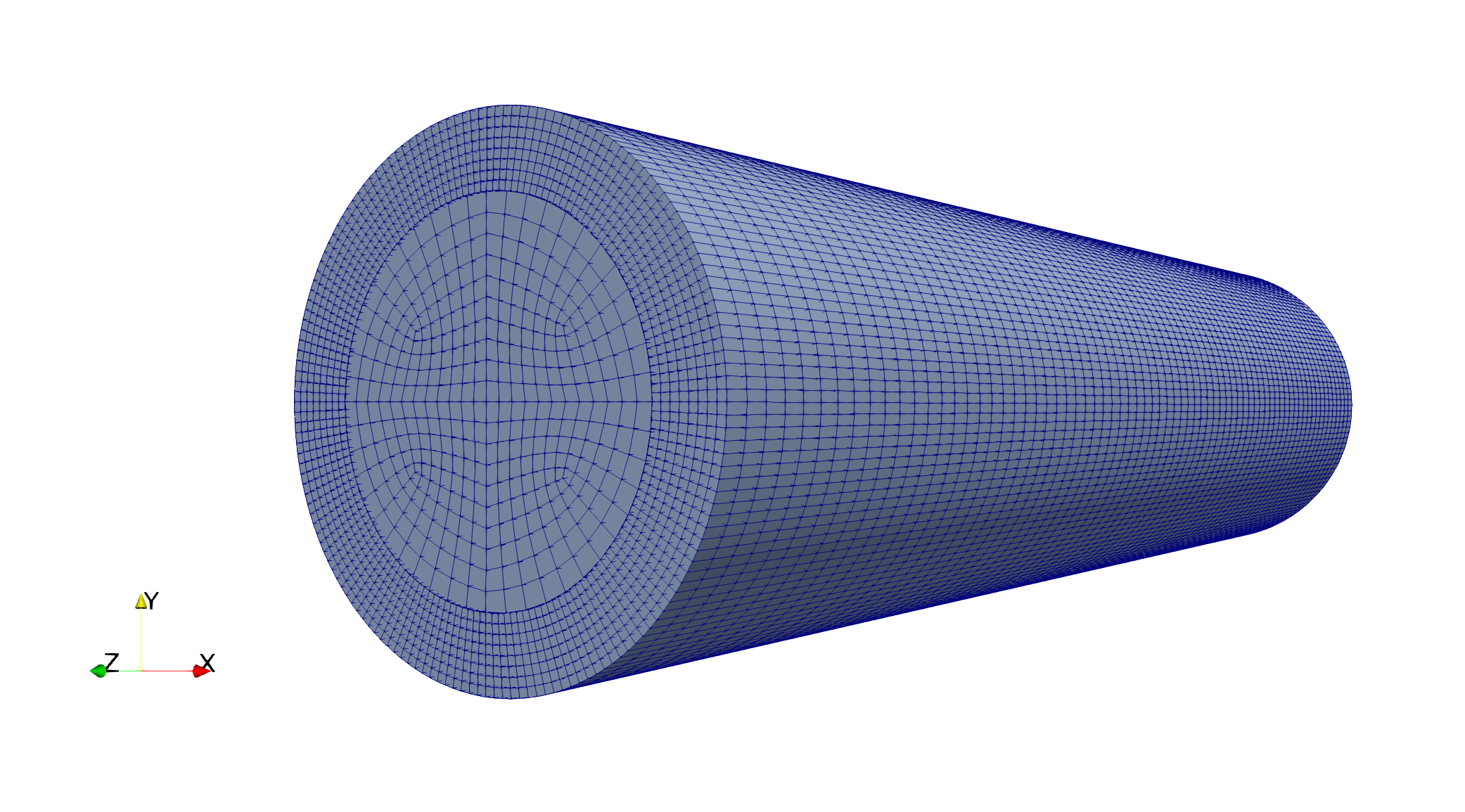}
		\subcaption{Start configuration of the stent model problem with fluid at rest}
		\label{Fig:stent_start}
	\end{subfigure}
	\caption{Stent model problem}
	\label{Fig:stent_setup}
\end{figure}

In Section \ref{sec:intro}, the motivation of capturing phenomena related to stenting  procedures through a bottom-up modeling approach was already mentioned. In the context of the direct interaction of blood flow and stent structure, two major complications determining the long-term outcome of the stenting procedure arise: Firstly, the influence of the FSI force on the onset of stent migration. The complex phenomenon of stent migration depends on a delicate interplay of the blood-stent interaction, stent-vessel wall frictional contact and in the long term even on growth and remodeling of the vessel wall itself and can thus not yet be reasonably captured by the proposed model. One of the involved factors, the growth and remodeling of the vessel wall, leads to the second major complication in the wake of stenting procedures: in-stent restenosis. Restenosis is currently thought to be connected to trauma of the vessel wall during the intervention itself, but also to a change in the flow pattern through the affected artery leading to oscillating wall shear stress (WSS) \cite{Auricchio2013, liang2009, holzapfel2005}. The additional roughness introduced by the stent itself has an effect on the latter. This motivates the following example of an immersed stent geometry inspired by material properties reported in \cite{Holzapfel2010} for the Taxus Libert\'{e}, which is used for stenting of coronary arteries. Nevertheless, since the beam constitutive model is given by a hyperelastic material model, not taking into account plasticity or pre-stressing, these aforementioned material properties have been adapted such that the fully expanded stent will touch the fluid boundary.

\begin{figure}[h!]
	\begin{center}
		\begin{subfigure}{0.45\textwidth}
			\includegraphics[width=\textwidth]{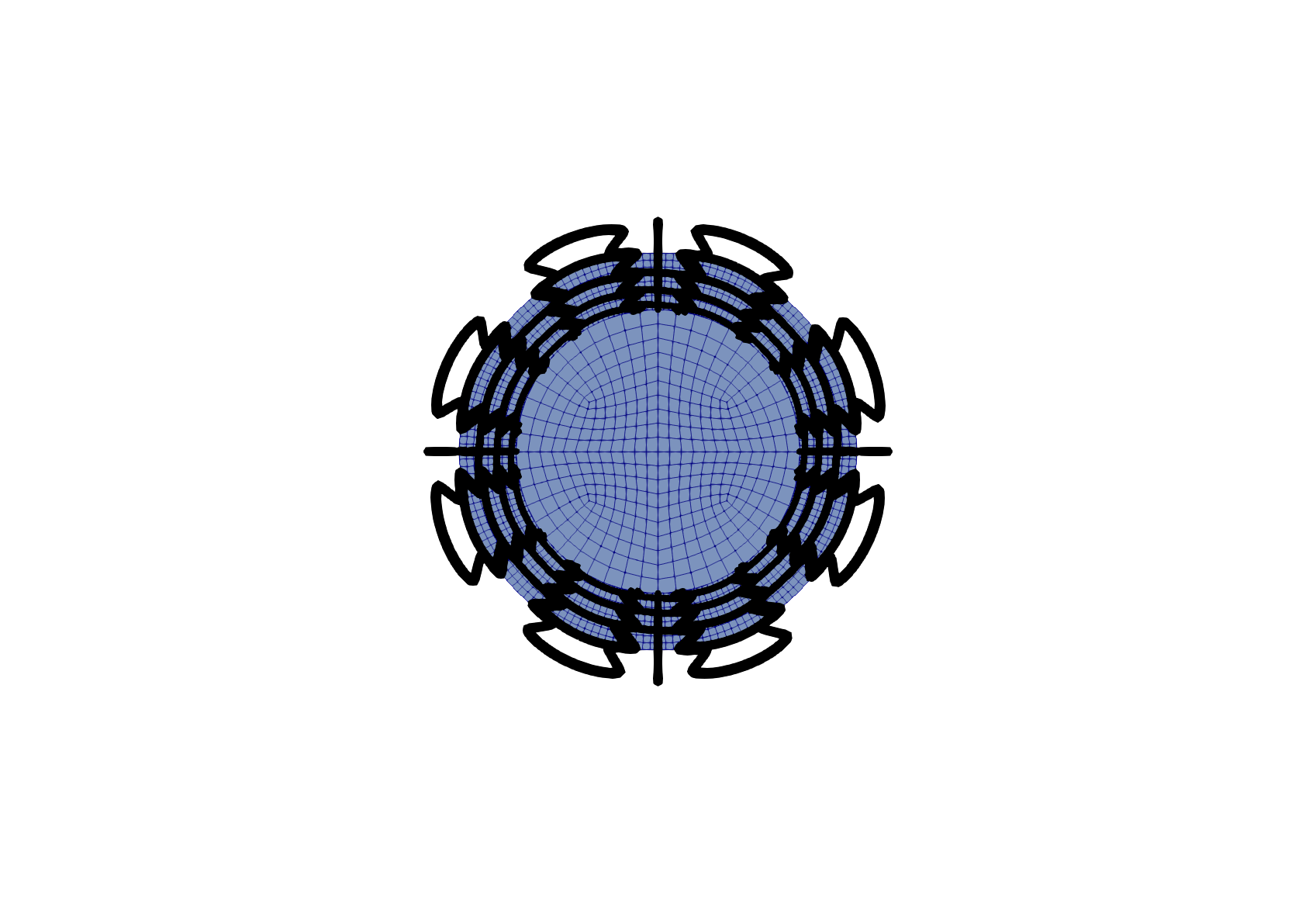}
			\subcaption{Visualization of the initial position of the stent within the fluid pipe}
			\label{subfig:outflow_initial}
		\end{subfigure}
    \hspace{1cm}
		\begin{subfigure}{0.45\textwidth}
			\includegraphics[width=\textwidth]{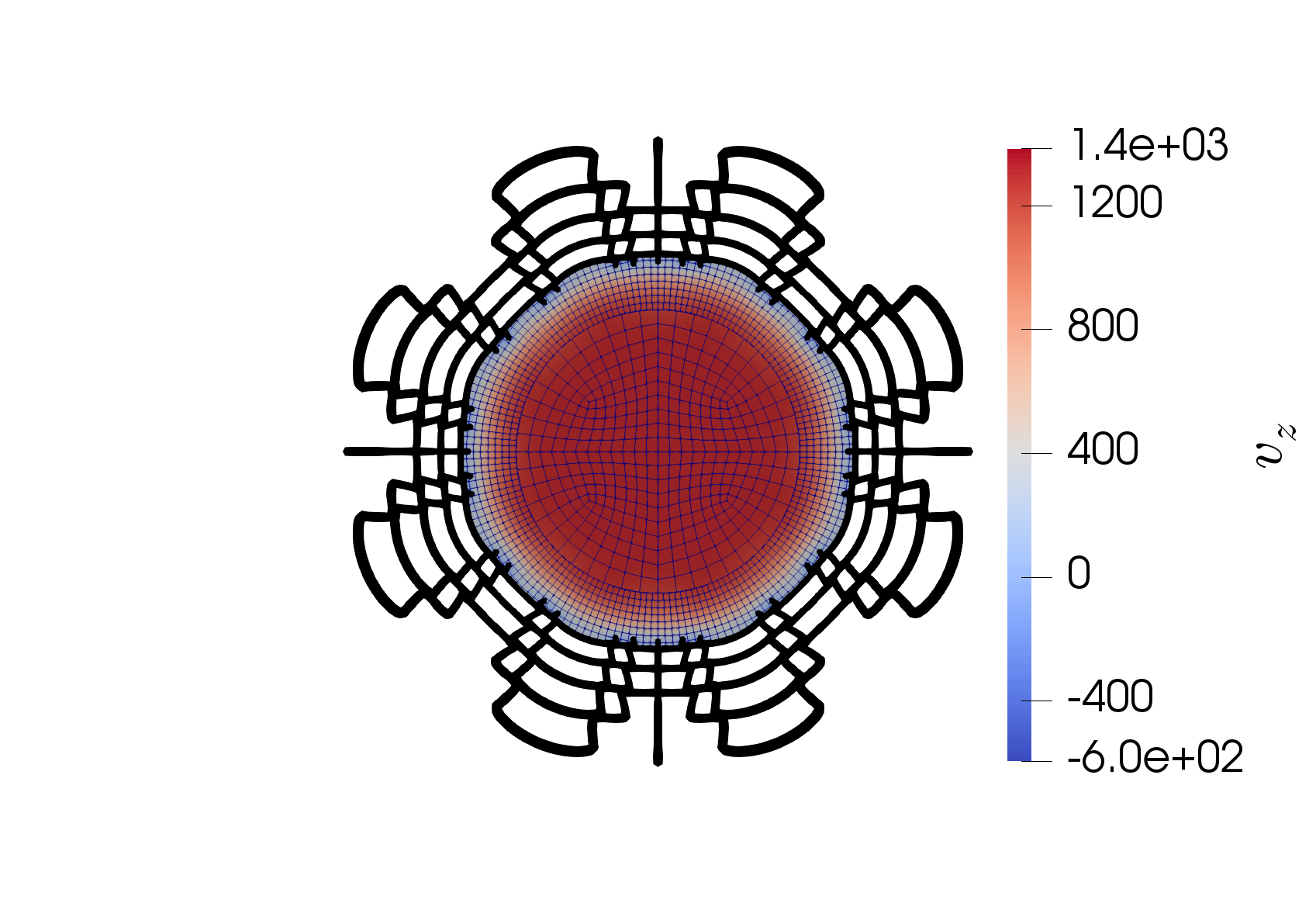}
			\subcaption{Effect of the stent on the fluid flow in axial direction}
			\label{subfig:outflow_open}
		\end{subfigure}
	\end{center}
	\caption{Axial view of the fluid flow within the pipe in $\dfrac{\mathrm{mm}}{\mathrm{s}}$}
	\label{fig:outflowsurf}
\end{figure}

The stent geometry itself, depicted in Figure \ref{Fig:stent_geometry}, is based on \cite{Auricchio2013}, has an initial radius of $1.25\:\mathrm{mm}$, a length of $13\:\mathrm{mm}$, and is made up of 3488 Simo-Reissner beam elements. The beams have a Young's modulus of $E=6.2\cdot 10^{10}\: \dfrac{\mathrm{g}}{\mathrm{mm}\cdot \mathrm{s}^2}$, a Poisson ratio $\nu=0.3$, and a radius of $r=0.03\:\mathrm{mm}$. Figure \ref{Fig:stent_setup} shows the setup of the model problem, for which the stent is half-way immersed in the fluid domain. The fluid domain consists of $113,800$ fluid elements, has a radius of $1.75\:\mathrm{mm}$, and a length of $15\:\mathrm{mm}$. The density is set to $\rho = 0.001\:\dfrac{\mathrm{g}}{\mathrm{mm}}$, the dynamic viscosity $\gamma=0.003\:\dfrac{\mathrm{g}}{\mathrm{mm} \cdot \mathrm{s}}$, and the time step size $\Delta t=0.001\:\mathrm{s}$ is used. The $113,800$ fluid elements are further subdivided into two independent meshes: a coarser inner cylinder mesh with a radius of $1.25\:\mathrm{mm}$ and a finer outer layer as visualized in Figure \ref{Fig:stent_start}. Both fluid meshes are then coupled using the mortar finite element method for surface coupling with condensed dual Lagrange multiplier shape functions as introduced in \cite{ehrl2014}. This mesh tying problem within the fluid field represents a rather complex application. This is to show that it is not only possible to include highly nonlinear phenomena within the structure field as demonstrated in Section \ref{subsec:contact}, but that the flexibility of the proposed approach also allows for complex models of the fluid field. For the application at hand, this is especially beneficial, since it is a priori known that the stent geometry will only move through the outer fluid layer, and it is expected that its effect on the fluid flow will also be restricted to this part of the fluid domain. Figure \ref{subfig:outflow_initial} shows the initial position of the stent at the edge of the outer fluid mesh layer.

To model the extension of the stent by the balloon, the internal pressure is successively increased from $t=0\:\mathrm{s}$ to $t=0.1\:\mathrm{s}$ using the formula

\begin{equation}
\left(\begin{matrix}
\bar{p}\cdot x\cdot t \\
\bar{p}\cdot y\cdot t
\end{matrix}\right).
\end{equation}

For a fully expanded stent geometry, as shown in Figure \ref{subfig:outflow_open}, $\bar{p}=3.1\cdot10^3\: \dfrac{\mathrm{g}}{\mathrm{mm}\cdot \mathrm{s}^2}$ is used. For simplicity, the fluid is assumed to be at rest and oblivious to the stent for the entire expansion procedure and the structure is simulated quasi-statically, thus neglecting its inertia. Once the stent is in place, the spatially constant fluid inflow velocity prescribed at the inlet boundary to the right is ramped up during $50$ time steps to a maximum velocity of $1000\:\dfrac{\mathrm{mm}}{\mathrm{s}}$, while no penetration, free slip boundary conditions are enforced on the cylinder barrel of the fluid domain.

\begin{figure}[h!]
	\includegraphics[width=\textwidth]{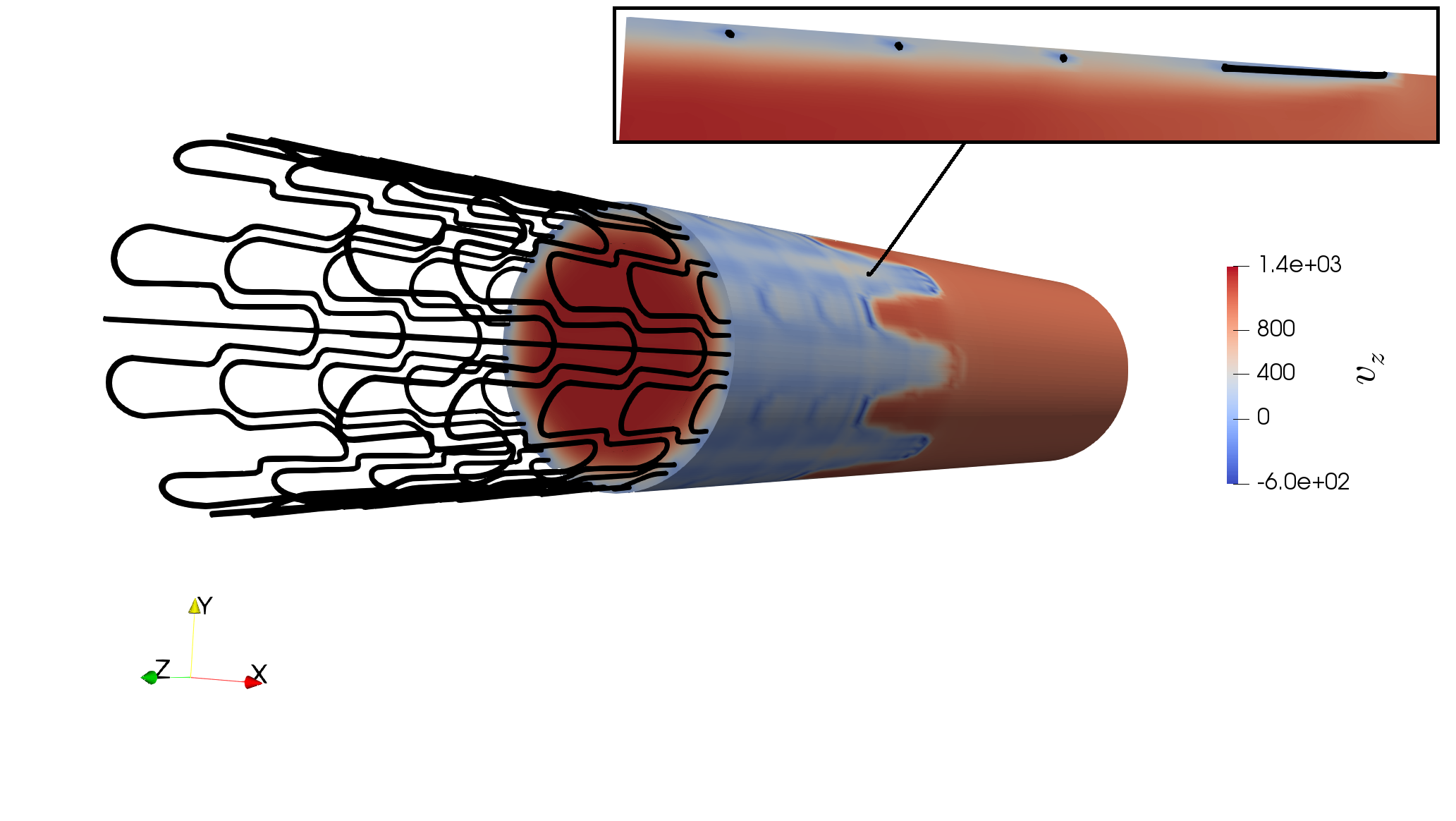}
	\caption{Final configuration of the stent model problem showing the axial velocity in $\dfrac{\mathrm{mm}}{\mathrm{s}}$ and in a slice through the origin at a 45$^\circ$ angle}
	\label{Fig:stent_end}
\end{figure}

Figure \ref{Fig:stent_end} shows that the current method can capture the roughness introduced by the complex fully expanded stent geometry while taking into account material properties and expansion behavior of the structure. It is noteworthy that the exact location of stent wires/struts and, thus, the location of the imposed Dirichlet coupling conditions depends on the solution of the nonlinear beam problem modeling the stent and is thus not known a priori as in the other examples.
To this end, Figure \ref{Fig:stent_end} demonstrates that features of the expanded stent geometry translate directly to the fluid flow. The fluid flow far from the stent is unaffected, as expected. On closer inspection of the slice through the origin at a 45$^\circ$ angle, backflow appears in the vicinity of the stent struts (colored in black), while the fluid is partly slowed down in between the cross beams. Also, the expansion procedure leads to larger displacements at the outskirts of the stent system, whereas the stabilizers and cross struts lead to stiffer a behavior far from the axial edges. This effect leads to even more flow back of the fluid near the upstream edge of the stent, while a boundary layer of decreased velocity is created downstream, where the stent geometry displays a smaller radial displacement.

Note that this model problem still represents a considerable simplification of the complex interactions present in a stented artery. In particular the neglect of any vessel wall and according boundary conditions poses an immense change in the dynamics of the system model. Nevertheless, the result shall serve as a proof of concept that the effect of established stent models, using one-dimensional beam elements as proposed in \cite{zunino2016, tambavca2010mathematical, Tambaca2010}, can be captured by our FBI scheme. This represents a sound starting point for extensions to more complex models by including surface-coupled fluid-structure interaction \cite{Kloppel2012, mayr2014} to model the influence of the vessel wall, as well as solid-beam interaction such as in \cite{steinbrecher2020} in order to model interactions of the stent with the vessel wall.

\section{Conclusions}
\label{sec:conclusions}

We have proposed an embedded finite element method to submerse one-dimensional beam elements in a three-dimensional fluid mesh using a GPTS-type coupling discretization. The introduced kinematic constraint is enforced via a penalty regularization directly on the beam centerline, thus making it a truly mixed-dimensional 1D-3D coupling problem as well as a highly efficient modeling strategy for immersed slender structures regarding computational complexity as well as mesh creation. This methodology was implemented for one-way coupling of the fluid to the beam and vice versa introducing all relevant components for a novel weak Dirichlet-Neumann partitioned algorithm. For both cases, the convergence of the resulting system solution with respect to the penalty parameter as well as the robustness of the segmentation algorithm to arbitrary positions within the background mesh was shown. The spatial convergence behavior of the fluid solution computed with the proposed 1D-3D coupling approach  under uniform mesh refinement as compared with a finely-resolved 3D CFD solution was shown and the approach, thus, validated. The general possibility of modeling highly complex nonlinear problems such as contact as well as mesh tying within the structure and/or fluid domain, and thus the flexibility of the proposed approach, was demonstrated by selected examples. In addition, limitations of the proposed one-way coupling strategies such as the exact capturing of interface phenomena have been pointed out. A simplified motivational example for possible practical application in form of an immersed stent geometry has been given as proof of concept for our modeling approach, and the additional nonlinearity introduced by the geometry change of the stent was well captured and visualized.

Obvious further research relates to the extension of the shown one-way coupling schemes to a stable two-way coupling algorithm by applying sophisticated convergence acceleration techniques to deal with well-known challenges in the context of the added mass effect. From a more application-centered view, the extension of the proposed method to include classical surface-based FSI, i.e. to model blood-vessel wall interactions in addition to the demonstrated blood-stent interaction, is an additional topic of interest.

\vspace{1cm}
\bibliographystyle{elsarticle-num}
\bibliography{references}

\end{document}